\documentclass[%
reprint,
superscriptaddress,
nofootinbib,
amsmath,
amssymb,
aps,
prd,
]{revtex4-2}

\usepackage{graphicx}
\usepackage{dcolumn}
\usepackage{bm}
\usepackage{appendix}
\usepackage{ulem}

\usepackage{hyperref}
\usepackage{multirow}

\usepackage[T1]{fontenc} 

\usepackage{dcolumn}
    \newcolumntype{d}[1]{D{.}{\cdot}{#1}}
    \newcolumntype{.}{D{.}{.}{-2}}
    \newcolumntype{,}{D{,}{,}{-3}}

\usepackage{booktabs}
\usepackage{xcolor}
\usepackage{physics}

\newcommand{\aref}[1]{\hyperref[#1]{Appendix~\ref*{#1}}}

\begin{document}


\title{Robust constraints on tensor perturbations from cosmological data: a comparative analysis from Bayesian and frequentist perspectives}

\author{Giacomo Galloni}
\thanks{Corresponding author}
\email{giacomo.galloni@unife.it}
\affiliation{Dipartimento di Fisica, Universit\`a di Roma Tor Vergata, Via della Ricerca Scientifica, 1, 00133, Roma, Italy}
\affiliation{Dipartimento di Fisica, Universit\`a La Sapienza, P.le A. Moro 2, 00185, Roma, Italy}
\affiliation{Istituto Nazionale di Fisica Nucleare (INFN), Sezione di Roma 2, Via della Ricerca Scientifica, 1, 00133 Roma, Italy}
\affiliation{Dipartimento di Fisica e Scienza della Terra, Universit\`a degli Studi di Ferrara, via Giuseppe Saragat 1, 44122 Ferrara, Italy}

\author{Sophie Henrot-Versill\'e}

\author{Matthieu Tristram}

\affiliation{Université Paris-Saclay, CNRS/IN2P3, IJCLab, 91405 Orsay, France}

\date{\today}

\begin{abstract}
We analyze primordial tensor perturbations using the latest cosmic microwave background and gravitational waves data, focusing on the tensor-to-scalar ratio, $r$, and the tensor spectral tilt, $n_t$. Utilizing data from Planck PR4, BICEP/Keck, and LIGO-Virgo-KAGRA, we employ both Bayesian and frequentist methods to provide robust constraints on these parameters. Our results indicate more conservative upper limits for $r$ with profile likelihoods compared to Bayesian credible intervals, highlighting the influence of prior selection and volume effects. The profile likelihood for $n_t$ shows that the current data do not provide sufficient information to derive quantitative bounds, unless extra assumptions on $r$ are used. Additionally, we conduct a 2D profile likelihood analysis of $r$ and $n_t$, indicating a closer agreement between both statistical methods for the largest values of $r$. This study not only updates our understanding of the tensor perturbations but also highlights the importance of employing both statistical methods to explore less constrained parameters, crucial for future explorations in cosmology.

\end{abstract}

\maketitle

\section{Introduction}\label{sec: intro}

The Cosmic Microwave Background (CMB) is a relic from the distant past, containing within its faint signals the secrets of the Universe's beginnings \cite{Seljak_1997, Kamionkowski_1997, Kamionkowski_2016}. Of particular interest is B-mode polarization, which detection would reveal the presence of primordial Gravitational Waves (GWs), predicted by inflationary scenarios \cite{Guzzetti:2016mkm, Seljak:1996ti, Hu:1997hv}.

For this reason, observing B-modes has been the goal of past, present, and future experiments such as the BICEP/Keck Array~\cite{Ade_2014}, the Simons Observatory~\cite{Ade_2019}, CMB-S4~\cite{abazajian2016cmbs4}, and the Light satellite for the study of B-mode polarization and Inflation from cosmic microwave background Radiation Detection (LiteBIRD)~\cite{Hazumi:2019lys, PTEP_LiteBIRD}. On top of that, \textit{Planck} represents the cornerstone of modern Cosmology \cite{Planck_parameters, Planck_like} and has recently provided the most precise constraints on the standard $\Lambda$CDM model with its fourth release (PR4) \cite{Rosenberg:2022, Tristram:2023}. In this last iteration, PR4 also provides information on B-modes, which, in combination with the measurements of BICEP/Keck, represents our most advanced knowledge on primordial GWs exploiting CMB data \cite{Tristram:2022,Galloni:2023}.

In parallel, GW interferometers are also proving to be a very insightful tool for providing complementary information on the primordial power spectrum of tensor modes \cite{Planck_2018}. Indeed, CMB experiments are notably not very effective in constraining its tilt ($n_t$). This is due to the lensing signal dominating the B-mode spectrum for $\ell > 300$. Although GW interferometers, such as LIGO-Virgo-KAGRA (LVK) \cite{aLIGO, aVirgo, ligo-virgo}, are designed to detect ripples in spacetime emanating from cataclysmic cosmic events, they also provide a unique synergy with CMB observations. Their ability to probe the GW energy density in a frequency range totally different from that of the CMB enhances our chances of constraining the primordial Universe scenarios \cite{Abbott_2019_ligo, ligo_meas}.

The primordial scalar and tensor perturbations are customarily parameterized with power laws as
\begin{equation}
    P_s(k) = A_s \qty(\frac{k}{k_s})^{n_s - 1} \qq{and} P_t(k) = A_t \qty(\frac{k}{k_t})^{n_t}\ ,
    \label{eq:power_laws}
\end{equation}
where $A_s$ ($A_t$), $n_s$ ($n_t$) and $k_s$ ($k_t$) are respectively the scalar (tensor) amplitude, spectral tilt and pivot scale. On top of this, the tensor-to-scalar ratio at a generic scale is defined as
\begin{equation}
    r_{k} = \frac{P_t(k)}{P_s(k)}\ ,
    \label{eq:r_definition}
\end{equation}
which, together with $n_t$, represent the tensor sector of parameter space. The state-of-the-art knowledge on this sector can be divided into two cases: in the first, the spectral tilt is fixed to the so-called consistency relation of single-field slow-roll inflation $n_t = -r_{0.05}/8$ \cite{Guzzetti:2016mkm}, and only $r_{0.05}$ is constrained through the data. In this context, the current lowest limit on $r_{0.05}$ is $r_{0.05} < 0.032$ at 95\% Confidence Level (CL) based on the combination of Planck and BICEP/Keck data~\cite{Tristram:2022}. On the other hand, one can avoid the choice of single-field slow-roll inflation violating the relation just mentioned. In this case, the current bounds are $r_{0.01} < 0.028$ and $-1.37 < n_t < 0.42$ at 95\% CL, accounting also for LVK data \cite{Galloni:2023}.

In a scenario where we do not have a detection of primordial GWs, the choice of the statistical analysis method used to extract information from the data is crucial. Bayesian methods, based on Markov Chain Monte Carlo (MCMC), have been pivotal \cite{MCMethod, Ulam, MonteCarlo, Lewis:2002ah, Lewis:2013hha}. However, challenges arise, particularly in dealing with volume effects that can mislead our conclusions in multidimensional parameter spaces \cite{Efstathiou:2023fbn}. This is particularly the case for the tensor sector as we shall see in the rest of the paper. Indeed, for a given sensitivity, the constraints on $n_t$ can be very broad if $r$ is pushed to very low values.

This work provides an update on the current knowledge on the parameters of the primordial tensor power spectrum using the latest \textit{Planck} release (PR4). After employing a MCMC approach to obtain information about the posterior probability, we gauge the impact of volume effects and prior choices by using the frequentist approach based on Profile Likelihoods (PLs), which offers independence from both \cite{Henrot_Versill__2015, Henrot-Versille:2016htt, Herold_2022, Murgia_2021, McDonough:2023qcu}. Indeed, MCMC and PL answer different and complementary questions, delivering us with a complete picture of the tensor parameter space.

In \autoref{sec: data}, we offer a concise overview of the datasets utilized in this study, encompassing contributions from the CMB derived from \textit{Planck} and BICEP/Keck, as well as insights from a GW standpoint, particularly from the LVK collaboration. 
\autoref{sec: methodology} discusses the methodologies underpinning both the MCMC and the PL analyses. Special attention is paid to the techniques employed to extract confidence intervals, ensuring their adherence to the correct statistical properties. Finally, \autoref{sec: Results} presents our findings, revisiting the outcomes of \cite{Galloni:2023} through the lens of updated datasets using the MCMC perspective. Simultaneously, we introduce a novel frequentist exploration of the tensor sector of parameter space using PL, shedding light on volume effects and prior-choice dependencies that impact Bayesian results.

\section{Datasets} \label{sec: data}

\subsection{Data}
\subsubsection{\textit{Planck}}

This study uses \textit{Planck}'s PR4 maps\footnote{Available on the \textit{Planck} Legacy Archive: \url{http://pla.esac.esa.int}.}. They were produced by the NPIPE processing pipeline, which reconstructs temperature- and polarization-calibrated frequency maps from Planck's LFI and HFI data.
The NPIPE process, detailed in \cite{planck2020-LVII}, incorporates data from previously neglected repointing periods, along with several enhancements that reduce noise and systematic errors across frequencies and component-separated maps, enhancing consistency between different frequencies. 

We use several likelihoods that cover the multipole range from $\ell=2$ to $\ell=2500$. For large angular scales in temperature ($\ell\leq 30$), we consider the Commander $TT$ likelihood (lowlT) based on a Bayesian posterior Gibbs sampling that combines the separation of astrophysical components and the estimation of likelihood \cite{Eriksen_2008, Planck_like}. For large angular scales in polarization, we use the Low-$\ell$ Likelihood Polarized for \textit{Planck} (LoLLiPoP) based on the Hamimeche-Lewis approximation for the $EE$, $BB$ and $EB$ power spectra \cite{Tristram:2022}, which can cover multipoles $\ell \leq 150$. The cross-correlation spectrum TE is not used at large scales.
At small angular scales ($\ell > 30$), we alternatively use the high-$\ell$ Likelihood Polarized for \textit{Planck} (HiLLiPoP) or CamSpec which both combine the $TT$, $TE$ and $EE$ CMB spectra over a large fraction of the sky (75\% and 80\% respectively). HiLLiPoP is a multi-frequency likelihood based on cross-spectra of the 100, 143 and 217\,GHz frequency maps, with astrophysical models for the residuals of foreground emissions \cite{Tristram:2023}. CamSpec is based on cross-spectra at 143 and 217\,GHz which are pre-processed by a cleaning procedure using the 545\,GHz maps as a template of Galactic dust emission \cite{Rosenberg:2022}. Each likelihood comes together with its own nuisance parameters mostly related to instrumental calibration and residual foreground modeling.

\subsubsection{BICEP/Keck Array}

We use the BICEP/Keck likelihood (BK18), representing data collected by the BICEP2, Keck Array and BICEP3 CMB polarization experiments up to the 2018 observing season~\cite{BICEP_2021}. This likelihood is based on the Hamimeche-Lewis approximation~\cite{Hamimeche_2008} for the joint likelihood of the BB auto- and cross-spectra obtained across multiple frequency maps: BICEP/Keck (two at 95 GHz, one each at 150 and 220 GHz), WMAP (23 and 33 GHz), and \textit{Planck} (PR4 at 30, 44, 143, 217, and 353 GHz). Covering an effective area of roughly 400 square degrees (equivalent to 1\% of the sky), this dataset is centered on a region characterized by minimal foreground emission. The data model encompasses Galactic dust and synchrotron emission, incorporating correlations between dust and synchrotron components.

\subsubsection{LIGO-Virgo-KAGRA interferometers}

We adopt the same approach as described in \cite{Galloni:2023, Planck_2018} taking advantage of the fact that GW interferometers probe scales ($k \sim 10^{16}$\,Mpc$^{-1}$) nearly 18 orders of magnitude above those probed by the CMB ($k \sim 10^{-2}$\,Mpc$^{-1}$). Under the assumption that the tensor tilt $n_t$ remains constant across the large frequency range between the CMB and GW interferometers, we use the upper limit on the energy density of GWs ($\Omega_{GW}$) provided by the LVK collaboration to obtain a constraint on $n_t$ at small scales, with \cite{Cabass_2016, Planck_2018} 
\begin{equation}
    \Omega_{GW}(k) = \frac{r_{0.05} A_s}{24z_{eq}} \left(\frac{k}{k_t} \right)^{n_t}\ ,
\end{equation}
where we considered the pivot scale to be $k_t = k_s = 0.05$ Mpc$^{-1}$ and $A_s$ is the scalar amplitude defined in \autoref{eq:power_laws}.
Then, we define a Gaussian likelihood on the energy density of GWs
\begin{equation}
    -2\log\qty(\mathcal{L}_{LVK}) = \frac{\qty(\Omega_{GW} - \mu_{LVK})^2}{\sigma_{LVK}^2}\ ,
\end{equation}
predicted as centered on $\mu_{LVK} = 0$ and with a $\sigma_{LVK}$ that equals half of the LVK's 95\% bound \cite{Planck_2018}.

In particular, we use the limit $\Omega_{GW}(25\,Hz) < 6.6 \times 10^{-9}$ (95\% CL) provided in \cite{ligo_meas} using data from the third observing run of Advanced LIGO and Advanced Virgo (O3) combined with upper limits of the previous runs O1 and O2.

\subsection{Combinations of data}\label{sec: likes}
Throughout the remainder of this paper, we refer to ``PLK20(CamSpec/HiLLiPoP)'' as the combination of \textit{Planck} likelihoods:
\begin{itemize}
    \item \textit{Planck} PR3 low-$\ell$ TT \cite{Eriksen_2008, Planck_like},
    \item \textit{Planck} PR4 LoLLiPoP \cite{Tristram:2022},
    \item \textit{Planck} PR4 lensing \cite{carron_2022},
    \item \textit{Planck} PR4 high-$\ell$ (CamSpec \cite{Rosenberg:2022} or HiLLiPoP \cite{Tristram:2023}),
\end{itemize}
We then add the two other datasets:
\begin{itemize}
    \item BICEP/Keck array 2018 (BK18) \cite{BICEP_2021},
    \item LIGO-Virgo-KAGRA 2021 (LVK21) \cite{ligo_meas}.
\end{itemize}
to form the two main combinations we study in this work, PLK20(CamSpec)+BK18+LVK21 and PLK20(HiLLiPoP)+BK18+LVK21.
Furthermore, we will consider the same combinations without LVK21 to emphasize the role of GW interferometers.

Note that we slightly modified the high-$\ell$ likelihoods (both HiLLiPoP and CamSpec) to avoid any correlation with the low-$\ell$ LoLLiPoP. Indeed, to maximize the information coming from $BB$, we use LoLLiPoP up to $\ell=150$ and consequently adjust the minimum multipole of high-$\ell$ likelihoods for $EE$ at $\ell_{\rm min}=151$. 

As previously done in \cite{Tristram:2022}, we neglect correlations between \textit{Planck} and BICEP/Keck datasets and simply multiply the likelihood distributions. This is justified because the BK18 spectra are estimated on 1\% of the sky, while the \textit{Planck} analysis is derived from at least 50\% of the sky.

\section{Methodology} \label{sec: methodology}

In this work, we explore both the tensor-to-scalar ratio $r$ and the tensor spectral tilt $n_t$. This allows us to test the consistency relation imposed by the single-field slow-roll inflationary model ($n_t = -r/8$ at first order), leaving instead $n_t$ free to vary. The only assumption we make is that the tensor power spectrum can still be described as a power law (see \autoref{eq:power_laws}). In a context where $n_t$ is fixed, the pivot is usually chosen to match the scalar pivot scale of $k_t = k_s = 0.05$  Mpc$^{-1}$. Instead, in our case the pivot is taken to be $k_t = 0.01$ Mpc$^{-1}$, since this scale is close to the decorrelation scale \cite{Planck_2018}. 

It is important to emphasize that we also fit for the 6 parameters of the $\Lambda$CDM cosmological model as well as all nuisance parameters associated to the likelihoods (see \autoref{sec: likes} and the references there-in for more details). For this reason, we refer to the model we study in this paper as $\Lambda$CDM + $r_{0.01}$ + $n_t$. Then, taking into account the nuisance parameters, the size of the parameter space considered is of the order of $\sim30$ dimensions.

To compute the angular power spectra of all the observables considered here, we employ \texttt{CAMB} \cite{Lewis:2013hha, Lewis_2000, Howlett_2012}.

Before discussing the details of the Bayesian and frequentist approaches to parameter estimation, it is essential to understand that these two methodologies address complementary questions. Bayesian credible intervals provide information about the probability that the true value of a parameter falls within a given interval. Conversely, frequentist confidence intervals inform us about the probability of obtaining the observed data. Thus, together they provide a complete picture on the parameter space under consideration, but they must be compared with care. In fact, for example, \cite{Henrot-Versille:2016htt,Couchot_2017,Campeti:2022vom} shows the importance of comparing and confronting results from both methods in order to gain a comprehensive understanding of the data and validate the robustness of the conclusions.

\subsection{MCMC analysis}

MCMC, a versatile and indispensable tool of Bayesian statistics, has become fundamental in the navigation of intricate landscapes of probability distributions \cite{MCMethod, Ulam, MonteCarlo, Lewis:2002ah, Lewis:2013hha}. Its role in exploring large parameter spaces becomes particularly apparent when seeking information about individual parameters. Indeed, such a procedure provides us with the marginalized posterior of a parameter given a set of data, which we can use to find the degree of belief of such a parameter. 

If a physical boundary is present (as in our case with $r_{0.01}$), it is naturally encoded in a Bayesian framework by the prior distribution. Indeed, considering a generic parameter named $\theta$ and a dataset $x$, the posterior $P(\theta|x)$ is obtained through the Bayes theorem, which reads
\begin{equation}
    P(\theta|x) = \frac{\mathcal{L}(x|\theta)\Pi(\theta)}{\mathcal{E}(x)}\ .
\end{equation}
Here, $\mathcal{L}(x|\theta)$ is the likelihood, $\Pi(\theta)$ is the prior and $\mathcal{E}(x)$ the evidence. A positivity bound on $\theta$ can be easily accounted for choosing
\begin{equation}
    \Pi(\theta) = 
    \begin{cases}
        0 \qquad \theta<0 \\
        1 \qquad \theta\geq0
    \end{cases}\ .
\end{equation}

All this comes without ambiguity in Bayesian statistics, as the random quantity evaluated here is $\theta$ and we are reconstructing its degree of belief.

Specifying the discussion to the tensor sector, \cite{Galloni:2023} reports a comprehensive analysis on the two most used approaches for $r_{0.01}$ and $n_t$. Here, we follow the same procedure chosen in that work: Single-Scale Approach (SSA). This implies the introduction of a lower cutoff in the tensor-to-scalar ratio of $r_{0.01} > 10^{-5}$. The underlying idea is that this amplitude level is far from detectable with current experiments, thus we cut the prior to avoid pathological behaviors of the MCMC. From a physical point of view, we know that some minimal B-mode signal is present even in single-field slow-roll inflation with no primordial tensor modes. Indeed, we have measured scalar perturbations of our Universe, which source GWs at second-order in the perturbations. In particular, those would produce a B-mode signal equivalent to $r \simeq 10^{-7}$ on large-scales and $r \simeq 10^{-5}$ on small-scales \cite{acquaviva2003SecondOrderCosmologicalPerturbationsInflation, matarrese1998RelativisticSecondorderPerturbationsEinsteinde, baumann2007GravitationalWaveSpectrumInduced, ananda2007CosmologicalGravitationalWaveBackground, fidler2014IntrinsicBmodePolarisationCosmic}.

Despite all this, in a Bayesian framework, volume effects can have an important role. In a multidimensional problem, if a large part of the probability volume is in a certain area, the final posterior will be drawn towards that region just as a result of the marginalization procedure. On top of this, excluding a priori parts of the parameter space (as we do, following the SSA prescription) could also bring to differences in the final results, highlighting the prior dependence of Bayesian statistics (see \cite{Galloni:2023}). In this context, another possible prior choice for a parameter whose order of magnitude is unknown is the log-uniform prior. This translates into equal weighting of the order of magnitude of this parameter \cite{Hergt:2021qlh}. 

These aspects can pose challenges in accurately gauging the significance of the obtained results and, possibly, mislead the derived conclusions. For this reason, in this work we confront and compare those results with the ones derived with the frequentist approach, i.e. the PL.

\subsection{Profile likelihood}\label{sec: metho_PL}

The procedure to get confidence intervals in a frequentist framework is slightly more involved due to the nature of frequentist intervals: in fact, the true values of parameters are not random variables, but rather fixed values that nature chooses. Thus, the boundaries obtained from a set of data are specific for the experiment considered and represent the random variable in this case. Repeating the experiment would cause these bounds to fluctuate. The ``coverage probability'' refers to the fraction of intervals that contain the true value of the parameters among the $N$ different repetitions of an experiment. The confidence intervals are then determined to have a coverage probability greater than or equal to a certain Confidence Level (CL) \cite{ParticleDataGroup:2022pth}.

To derive a confidence region that has the correct frequentist coverage properties, one can make use of likelihood ratio statistics. For multiparametric spaces, this amounts to constructing the PL: for fixed values of the parameter of interest (${\theta_i}$), we look for the maximum of the likelihood function in all the other dimensions (both for the physical parameters and the nuisance parameters). We then have access to the function $\chi^2_{min}(\theta_i)=-2 \ln {\mathcal{L}}_{max}(\theta_i)$, where $\mathcal{L}$ is the likelihood considered. The best fit (or $\min(\chi^2_{min}(\theta_i))$) gives the estimate of the parameter under consideration, which also corresponds to the maximum likelihood over all the other parameters. 

This procedure also ensures that its determination is independent of any change of variable $f(\theta_i)$, making it parameterization invariant. 

The error in the parameter $\theta$ can be deduced from the shape of the $\chi^2_{min}(\theta_i)$ function. For Gaussian distribution, the function is parabolic and the $1\sigma$ error bounds are simply obtained by a cut at $\Delta\chi^2_{\rm min} = \chi^2_{min}(\theta_i)-\min(\chi^2_{min}(\theta_i))=1$. When dealing with physical boundaries, one needs to use the Feldman-Cousins (FC) prescription and the Neyman construction \cite{ParticleDataGroup:2022pth, Feldman:1997qc}. Note that no integration in parameter space is performed to obtain the confidence intervals. Indeed, this is the manifestation of the core difference between Bayesian and frequentist approaches to parameter estimation mentioned above. The former requires to marginalize a posterior to extract bounds on a single parameter, while the latter does not.

One of the difficulties in building the PL is the precision with which we need to determine the values of $\chi^2_{min}(\theta_i)$. We must rely both on a very accurate minimizer and a boost to the accuracy parameters of the Boltzmann code \texttt{CAMB} \cite{planck2020-LVII}.

The PL procedure also generates a byproduct, the ``co-profiles''. In fact, we not only obtain a function $\chi^2_{min}(\theta_i)$, but also $N-1$ functions $\theta_j(\theta_i)$ where $j \neq i$. These co-profiles allow us to gauge the direction of degeneracy of other parameters on the profiled one and are very useful as a diagnostic tool of the PL procedure \cite{McDonough:2023qcu, Henrot-Versille:2016htt}. In other words, co-profiles allow us to explore the parameter space in the direction of the minimum $-2\log\mathcal{L}$ valley.

\subsection{Feldman-Cousins prescription}

The first step of the FC prescription is to assume a true value of the parameter of interest $\theta$. In fact, for each value of the parameter we can find an interval $\qty[x_1(\theta, \alpha), x_2(\theta, \alpha)]$ such that
\begin{equation}
    \alpha = \int_{x_1}^{x_2}\mathcal{L}(x|\theta)dx \ .
    \label{eq:neyman_coverage}
\end{equation}
Repeating this process for a set of ``true values'' of $\theta$ and drawing each segment $\qty[x_1, x_2]$ in a plot, we can find a ``confidence belt''. Now, having a measured value of $x = x_0$, the confidence interval $\qty[\theta_1(x), \theta_2(x)]$ is found by drawing a vertical line at $x_0$ and looking at the maximum and minimum values of $\theta$ (or its relative segment) that intersect this line.

Note that, actually, \autoref{eq:neyman_coverage} does not define uniquely $x_1$ and $x_2$, therefore, we need another equation to close the system. Here is where the FC prescription is defined. Consider the following test statistic
\begin{equation}
    \lambda(x, \theta) = \frac{\mathcal{L}(x|\theta)}{\mathcal{L}(x|\hat{\theta})}\ ,
\end{equation}
i.e. a likelihood ratio where $\hat{\theta}$ maximizes $\mathcal{L}(x|\theta)$. Solving \autoref{eq:neyman_coverage} while asking $\lambda(x_1, \theta) = \lambda(x_2, \theta)$ represents the FC construction.

This prescription not only recovers the correct coverage probability, but it is also able to naturally account for any eventual physical boundaries of the parameters, shifting ``automatically'' from a two-sided interval to a one-sided one. 
Suppose that
\begin{equation}
    \mathcal{L}(x|\theta) = \frac{1}{\sqrt{2\pi}}\exp\qty(-\frac{\qty(x-\theta)^2}{2})
\end{equation}
and that $\theta$ must be non-negative. Our physically allowed best estimate of $\theta$ is $\hat{\theta} = \max(0, x)$; then, we can write
\begin{equation}
    \lambda(x, \theta) = \frac{\mathcal{L}(x|\theta)}{\mathcal{L}(x|\hat{\theta})} = 
    \begin{cases}
        \exp\qty(-\frac{\qty(x-\theta)^2}{2}) &\qq{if} x\geq 0 \\
        \exp\qty(x\theta - \frac{\theta^2}{2}) &\qq{if} x<0
    \end{cases}\ .
\label{eq:FC_bound_prescription}
\end{equation}

Now, if the measured value of $x$ is too close to the physical limit or is negative, the correspondent vertical line would intercept only $x_1(\theta, \alpha)$, automatically defining an upper limit on $\theta$. At some point, increasing the measured value, we would instead intercept also $x_2(\theta, \alpha)$, shifting to a two-sided confidence interval. For this reason, the FC intervals are also said to be \textit{unified} (see \cite{Feldman:1997qc} for more details).

\section{Results}\label{sec: Results}

In this section, we fit for the parameters of the $\Lambda$CDM model extended to tensor perturbations with the dataset described in \autoref{sec: data} and the methodology presented in \autoref{sec: methodology}.

In particular, we focus on the tensor sector of the parameter space, i.e. $\qty{r_{0.01}, n_t}$. As regards the $\Lambda$CDM parameters, we do not find any significant deviation from the current state-of-the-art results \cite{Tristram:2023, Rosenberg:2022} (see \aref{app:LCDM}). We mention that we use a modified version of \texttt{Cobaya} \cite{torradoCobayaCodeBayesian2020} to obtain both Bayesian and frequentist results. For the latter, we added a new minimizer based on \texttt{MINUIT}\footnote{\url{https://github.com/CobayaSampler/cobaya/pull/332}.} \cite{James:1975dr}, which outperformed the two alternatives implemented in Cobaya (\texttt{py-BOBYQA} \cite{cartis2018ImprovingFlexibilityRobustnessModelBased} and \texttt{scipy} \cite{virtanen2020SciPyFundamentalAlgorithmsscientific}).

Various configurations of \texttt{CAMB} and \texttt{MINUIT} have been tested in terms of reliability to obtain the absolute maximum of the likelihoods. More details can be found in \aref{app:accuracy_pl}.

Note that very recently many new tools have been developed to perform PLs, which are not exploited in this work \cite{Nygaard_2023, Holm:2023uwa, Karwal:2024qpt}.

\subsection{MCMC analysis}\label{sec:pl_mcmc_res}

\begin{figure}[t]
    \centering
    \includegraphics[width=\hsize]{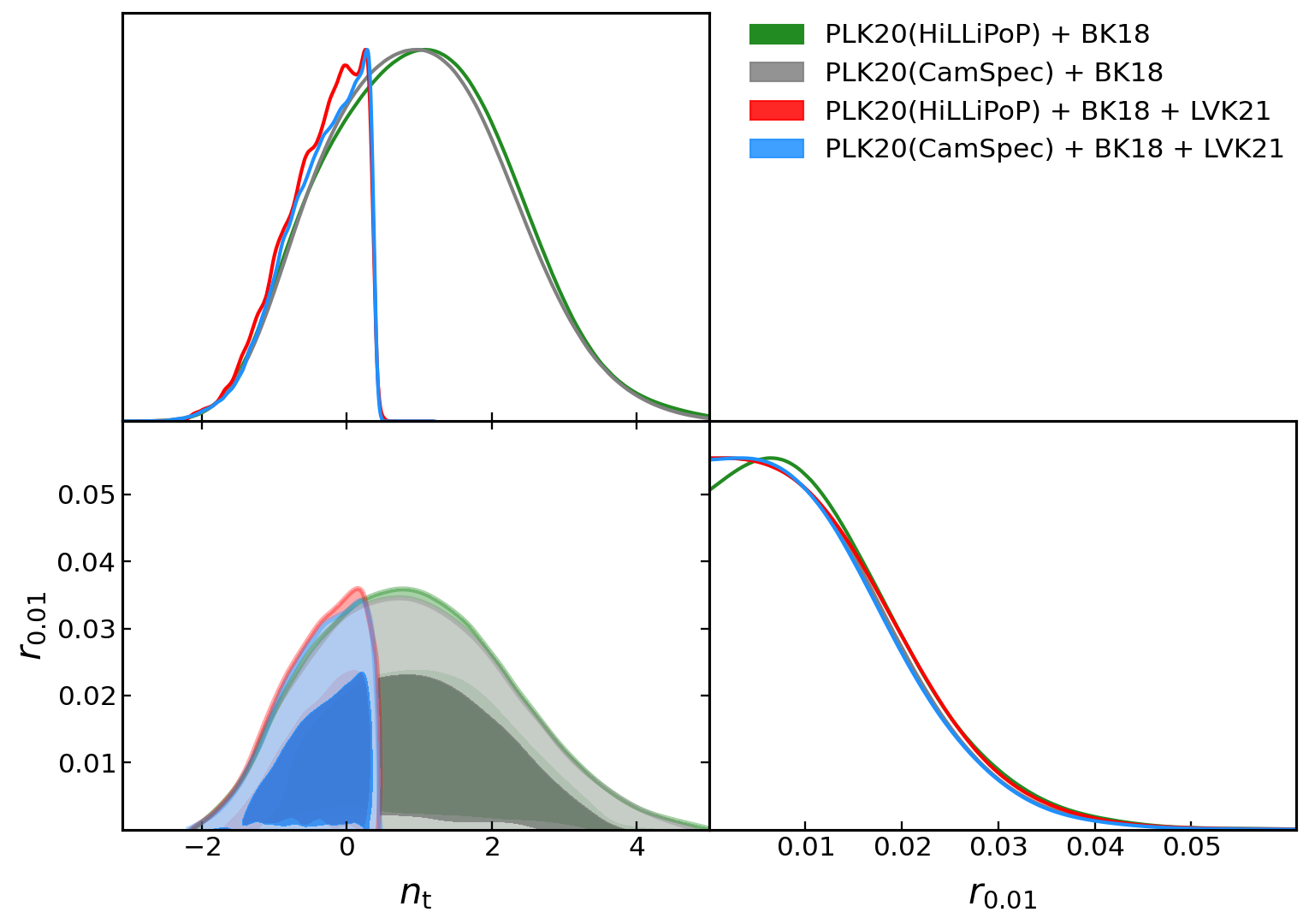}
    \caption{2D posterior of $r_{0.01}-n_t$ for \textit{baseline} + CamSpec and \textit{baseline} + HiLLiPoP.}
    \label{fig:mcmc_camspec_hlp}
\end{figure}

\autoref{fig:mcmc_camspec_hlp} shows the 2D posterior distributions for $r_{0.01}$ and $n_t$ from the MCMC analysis for the different datasets.

\begin{figure*}
    \centering
    \includegraphics[width =0.49\textwidth]{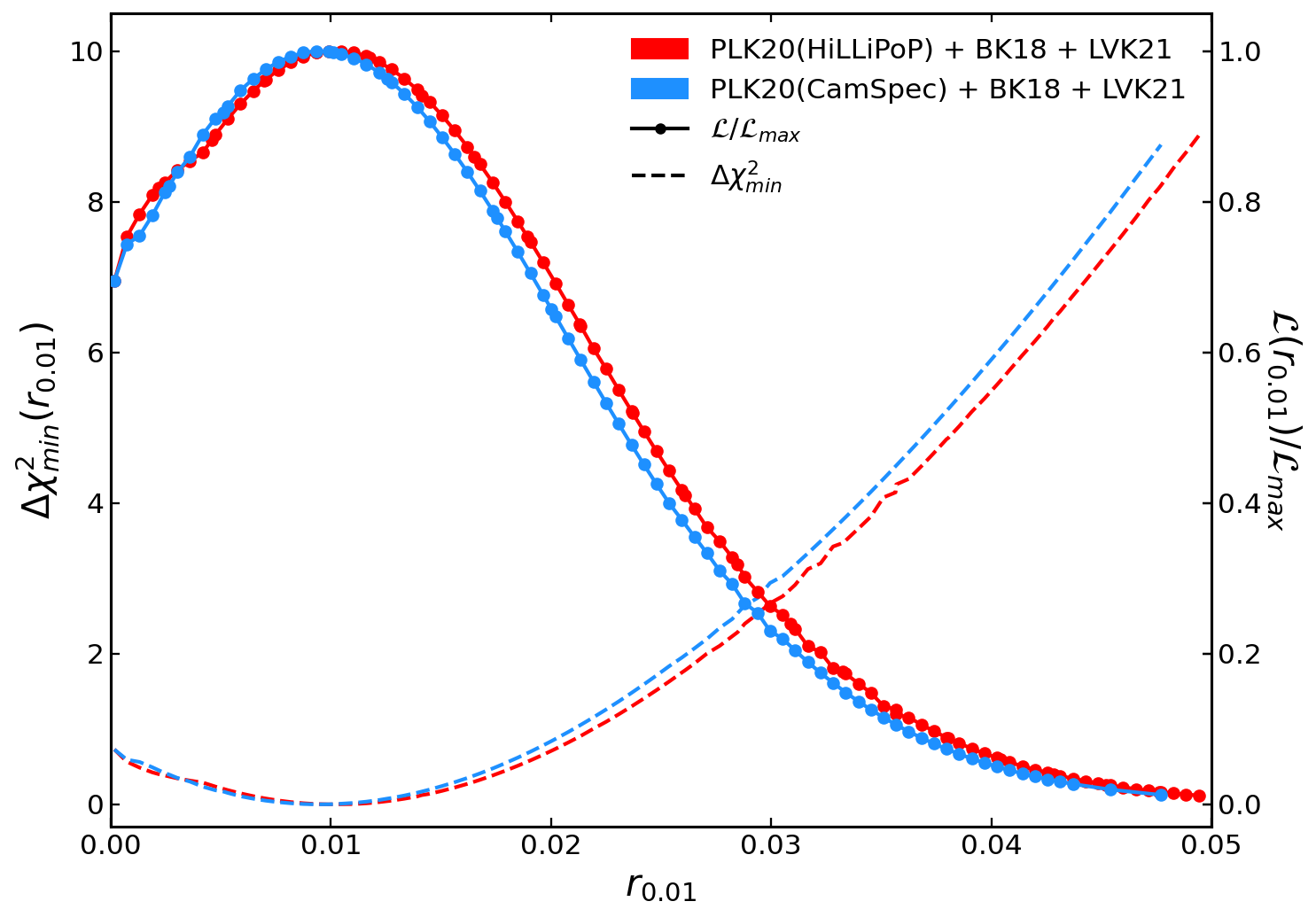}
    \hfill
    \includegraphics[width =0.49\textwidth]{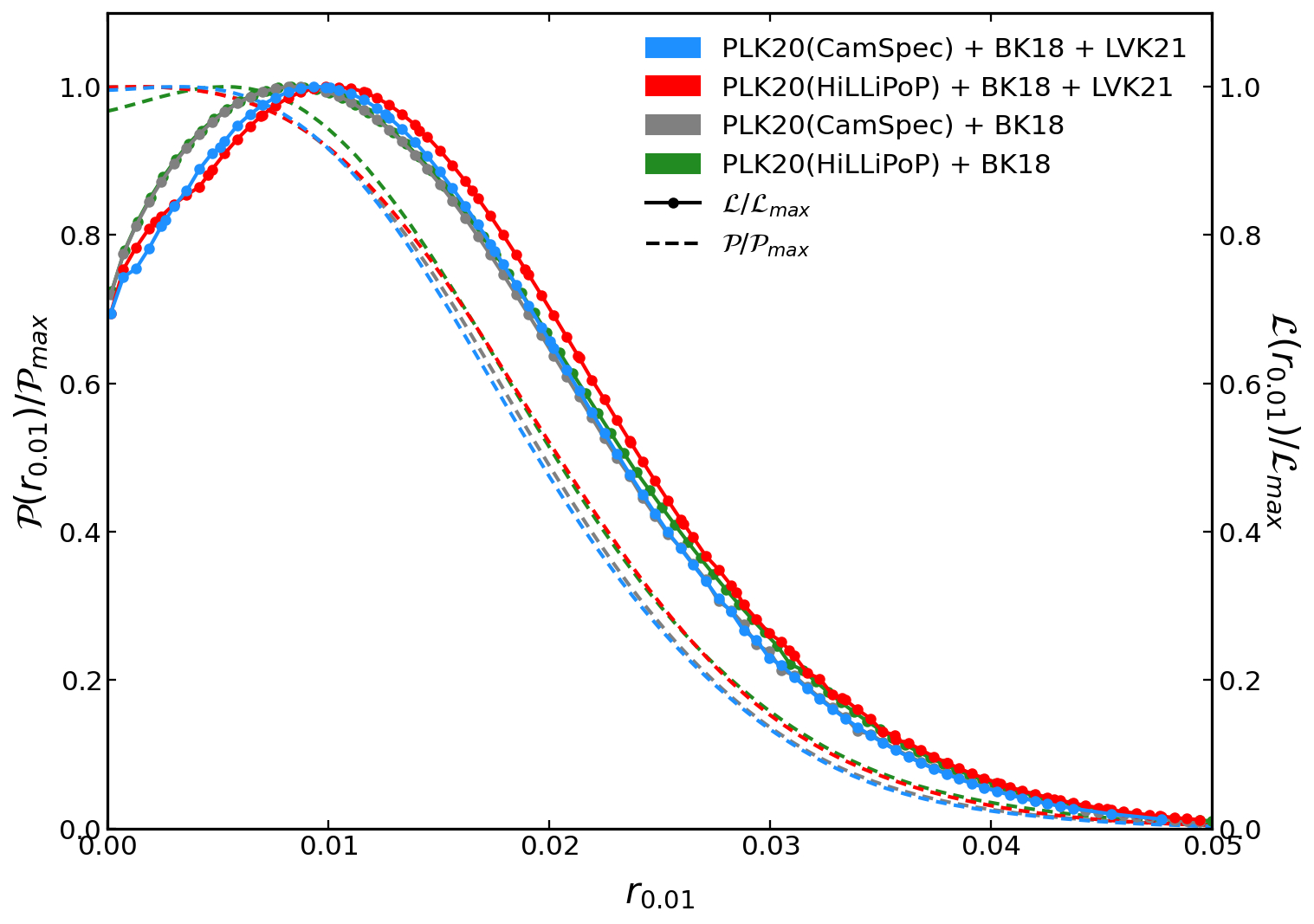}
    \caption{\textit{Left}: $\Delta\chi^2_{\rm min}(r_{0.01})$ and $\mathcal{L}(r_{0.01})/\mathcal{L}_{\rm max}$ for PLK20(CamSpec) + BK18 + LVK21 and PLK20(HiLLiPoP) + BK18 + LVK21. \textit{Right}: Profile likelihood ratio (solid lines) and posterior distributions (dashed lines) for $r_{0.01}$.}
    \label{fig:r_profiles}
\end{figure*}

We start by comparing the results between different choices on the high-$\ell$ part of CMB, i.e. CamSpec or HiLLiPoP. In particular, the 95\% CL marginalized intervals for PLK20(CamSpec) + BK18 + LVK21 are
\begin{equation}
    r_{0.01} < 0.028 \qq{and} -1.36<n_t<0.42\ ,
\end{equation}
while for PLK20(HiLLiPoP) + BK18 + LVK21
\begin{equation}
    r_{0.01} < 0.029 \qq{and} -1.39<n_t<0.41\ .
\end{equation}

These intervals are almost identical, suggesting that this choice does not significantly impact the tensor sector, as expected. 
Furthermore, these are also very similar to the results presented in \cite{Galloni:2023} which used a slightly different combination of datasets indicating the robustness of the derived limits regarding the choice of high-$\ell$ CMB likelihood \footnote{For completeness, \cite{Galloni:2023} used a combination of datasets made up of LoLLiPoP of PR4 + \textit{plik} and lensing from PR3 + BK18 + LVK21 (see \autoref{sec: data}).}. The bounds we obtain here for $n_t$ are typically broader than the ones reported in~\cite{Planck_2018}. We note that in~\cite{Planck_2018}, the tensor perturbations are characterized by fitting $r$ at two different scales which can be affected by prior effects as discussed in~\cite{Galloni:2023}.

As a test to gauge the importance of GWs interferometers, we also repeat the analysis removing LVK21. As expected, LVK21 severely constrains the tilt. For PLK20(CamSpec) + BK18 we obtain
\begin{equation}
    r_{0.01} < 0.029 \qq{and} -1.33<n_t<3.37\ ,
\end{equation}
while using PLK20(HiLLiPoP) + BK18 we obtain 
\begin{equation}
    r_{0.01} < 0.030 \qq{and} -1.35<n_t<3.40\ .
\end{equation}
Once again, these bounds are similar to the ones obtained in \cite{Galloni:2023}. The only sizable difference lies in the fact that the tensor spectral tilt has slightly shifted to higher values. We have investigated the causes of such a shift but have been unable to identify any specific behavior of the likelihood function, or distinctive features within the parameter space.

In conclusion of this section, we reassessed the state-of-the-art results using the most updated datasets (see \autoref{sec: data}). We found that the constraints on $r_{0.01}$ and $n_t$ have remained essentially stable as compared to \cite{Galloni:2023}, demonstrating that the constraints are driven mainly by the low-$\ell$ part of CMB polarization.

\subsection{Profile likelihood}

We have previously noted that results from the Bayesian analysis may be affected by volume effects or by the prior choice. In this section, we perform a frequentist analysis on the tensor sector of parameter space to gauge such effects.

\subsubsection{Tensor-to-scalar ratio} \label{sec:res_PL_r}

First, we calculate the PL on the tensor-to-scalar ratio. Our underlying model is $\Lambda$CDM + $r_{0.01}$ + $n_t$, so at each fixed value of $r_{0.01}$, the likelihoods are maximized both w.r.t. the standard $\Lambda$CDM parameters and $n_t$ (together with all the nuisance parameters in each likelihood).

The left panel of \autoref{fig:r_profiles} reports the results of PLK20(CamSpec) + BK18 + LVK21 and PLK20(HiLLiPoP) + BK18 + LVK21 in terms of both $\Delta\chi^2_{\rm min}(r_{0.01})$ and likelihood ratio $\mathcal{L}(r_{0.01})/\mathcal{L}_{\rm max}$ to emphasize the correspondence between these two quantities. On the other hand, the right panel shows the profile likelihood ratio for all the datasets defined above. All of them are very similar. However, one can note that HiLLiPoP results in slightly higher upper limit w.r.t. CamSpec. This resonates with the MCMC results of \autoref{sec:pl_mcmc_res}, where PLK20(HiLLiPoP) + BK18 + LVK21 gives slightly broader bounds. Indeed, even though the two profiles show quite similar widths, the HiLLiPoP case peaks at $r_{0.01} = 0.01$ while CamSpec at $r_{0.01} = 0.009$.

\autoref{fig:r_profiles} shows also the comparison between the posteriors from \autoref{sec:pl_mcmc_res} and PL distributions, suggesting that some volume effect is present in the Bayesian framework, pushing the posterior toward $r=0$. 

We emphasize that all PLs point to a best-fit value for the tensor-to-scalar ratio of $r_{0.01} \simeq 0.01$ while being completely consistent with a non-detection (i.e. $r_{0.01} = 0$). Instead, the maximum posterior tends to be around $r_{0.01} \simeq 0$ for every dataset (the only exception being PLK20(HiLLiPoP) + BK18, whose maximum posterior approaches the best-fit).

\subsubsection{Feldman-Cousins for \texorpdfstring{$r_{0.01}$}{TEXT}} \label{sec:FC_pl_r_res}

We now apply the FC prescription to recover the confidence interval with the correct coverage probability and accounting for the physical limit $r_{0.01} \geq 0$.

For a Gaussian distribution of data given a parameter $\theta$, we can include a physical bound in a frequentist framework by using the test statistic introduced by \autoref{eq:FC_bound_prescription}.

However, the PL ratios we obtain from both CamSpec and HiLLiPoP do not follow a Gaussian distribution far away from the best-fit value of $r_{0.01}$ (i.e. for $\Delta\chi^2 \gtrsim 2$). Indeed, we identify three distinct features in our PLs (see right panel of \autoref{fig:r_profiles}): all of them present a relatively large probability tail, PLK20(HiLLiPoP) + BK18 + LVK21 shows a bump at low values of $r_{0.01}$, and the likelihood ratio for the combinations without LVK21 seems to drop faster than a Gaussian on the left of the best-fit. We show in \aref{sec: generalized_fc} that accounting for these features with a more complex approach w.r.t. the one presented in \autoref{sec: metho_PL} makes no significant difference. 

\begin{figure}[t]
    \centering
    \includegraphics[width = 0.49\textwidth]{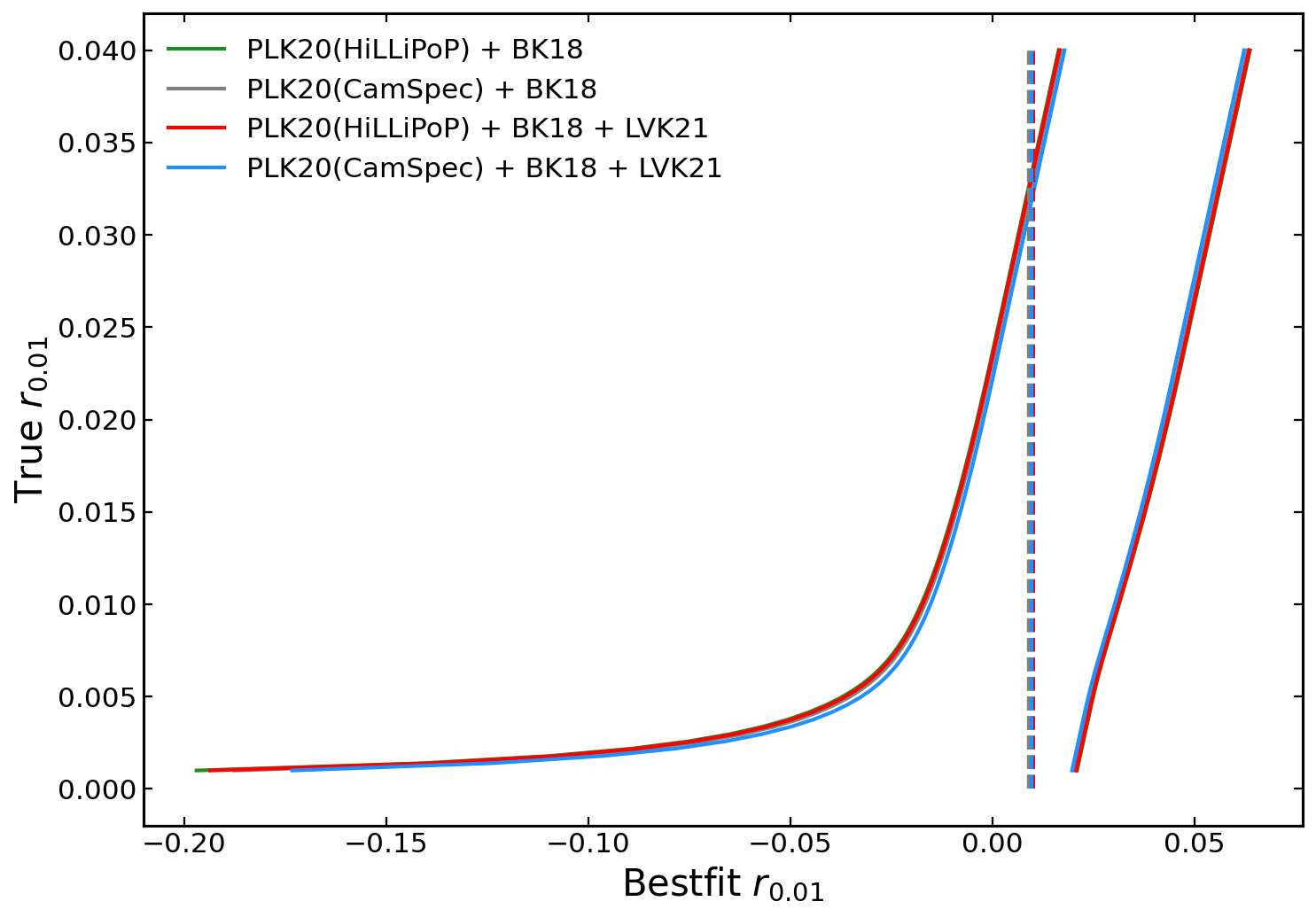}
    \caption{FC belts obtained including the positivity condition on $r_{0.01}$. The vertical dashed lines are the best-fit values obtained from each dataset.}
    \label{fig:camspec_fc}
\end{figure}

Finally, in \autoref{fig:camspec_fc} we report the results for all the datasets considered.
Intersecting these FC belts with the respective best-fit values, we obtain the upper bounds given in \autoref{tab:res_r_pl}.

None of these limits are affected by the volume effect, or the prior choice, previously discussed in \autoref{sec:res_PL_r}. As a result, the upper limits of the PL are more conservative than those derived with the Bayesian analysis (see \autoref{sec:pl_mcmc_res}). 

\begin{table}[t]
\small
\caption{\label{tab:res_r_pl}
95\% confidence intervals obtained with the FC prescription.}
\begin{ruledtabular}
\begin{tabular}{lc}
\textrm{Dataset}& \textrm{95\% Confidence Interval} \\
\colrule
PLK20(HiLLiPoP)+BK18 & $r_{0.01} < 0.033$  \\
PLK20(CamSpec)+BK18 & $r_{0.01} < 0.032$ \\ 
PLK20(HiLLiPoP)+BK18+LVK21 & $r_{0.01} < 0.033$ \\
PLK20(CamSpec)+BK18+LVK21 & $r_{0.01} < 0.032$ \\
\end{tabular}
\end{ruledtabular}
\end{table}

\subsubsection{Tensor spectral tilt} \label{sec:res_PL_nt}

\begin{figure}[t]
    \centering
    \includegraphics[width =0.49\textwidth]{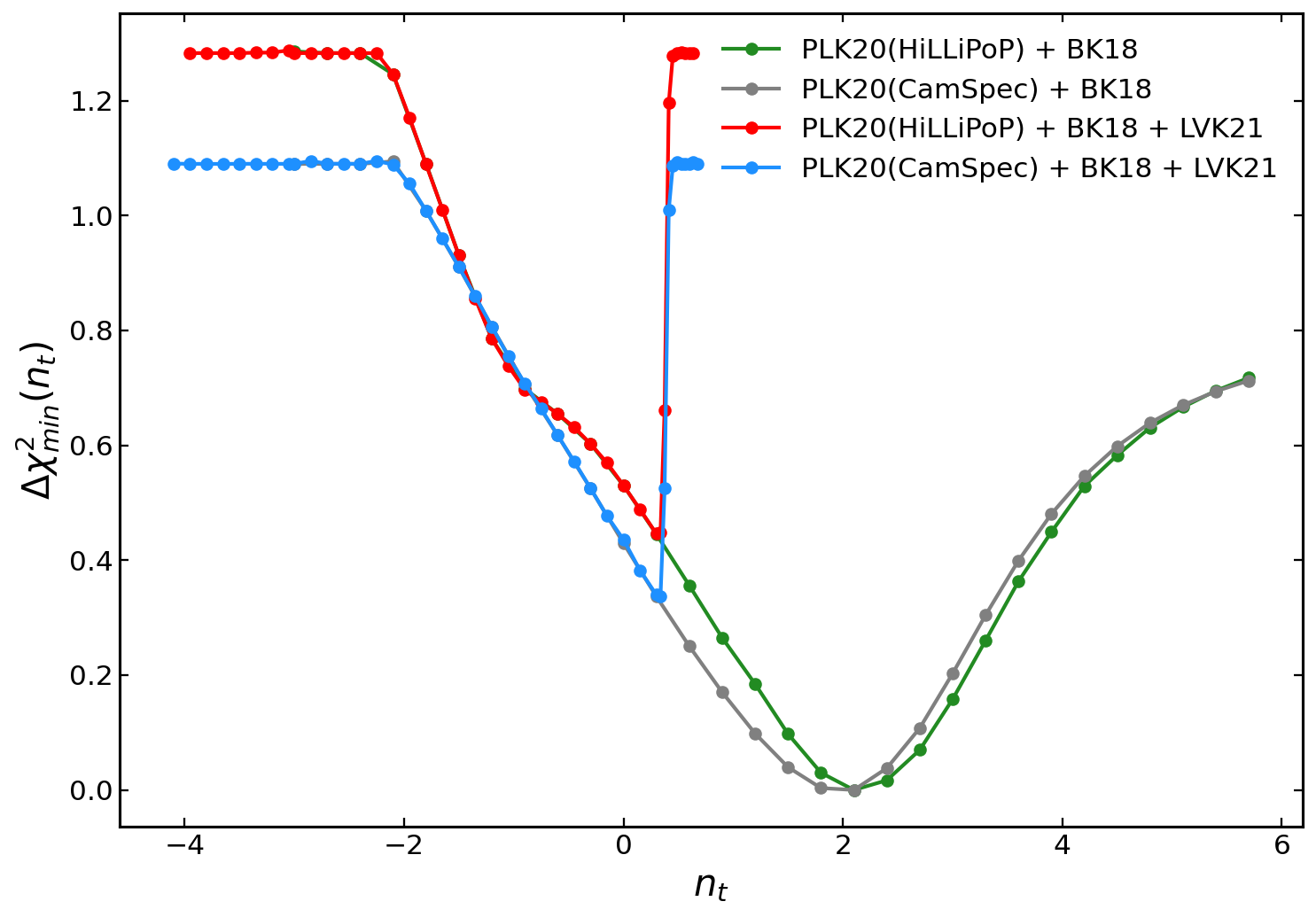}
    \caption{PL on the tensor spectral tilt. Both PLK(CamSpec) + BK18 + LVK21 and PLK(HiLLiPoP) + BK18 + LVK21 are normalized to the minimum of their counterpart without LVK21.} 
    \label{fig:nt_profiles}
\end{figure}

The profile likelihood on $n_t$ is shown on \autoref{fig:nt_profiles} for various combinations of the datasets.
Given the shape of the PLs, the derivation of quantitative upper bounds is beyond the scope of this work, and we shall nevertheless go to a qualitative description of the obtained results.

\begin{figure}[t]
    \centering
    \includegraphics[width =0.49\textwidth]{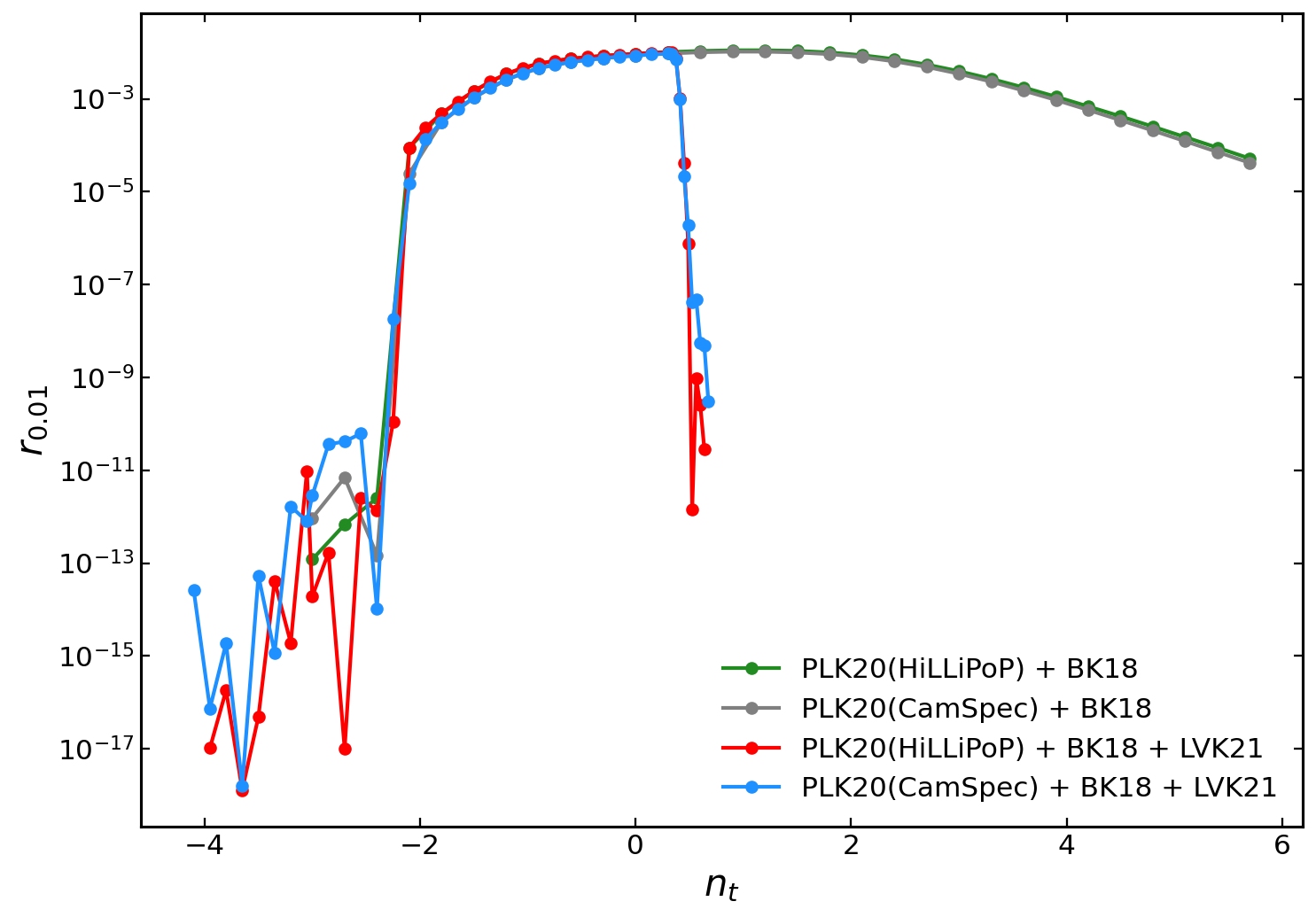}
    \caption{Co-profiles of $r_{0.01}$ as a function of the profiled value of $n_t$.}
    \label{fig:nt_coprofiles}
\end{figure}
First, PLs without GW data show a mild preference for a best fit around $n_t \simeq 2$. However, the fact that $\Delta\chi^2$ is always $\lesssim 1$ means that current data are not sufficiently constraining to statistically disentangle the different values of $n_t$. In fact, the part at low $n_t$ exhibits a constant-$\chi^2$ region, which we will call ``plateau'' and will be discussed below. 

\begin{figure*}[t]
    \centering
    \includegraphics[width =0.49\textwidth]{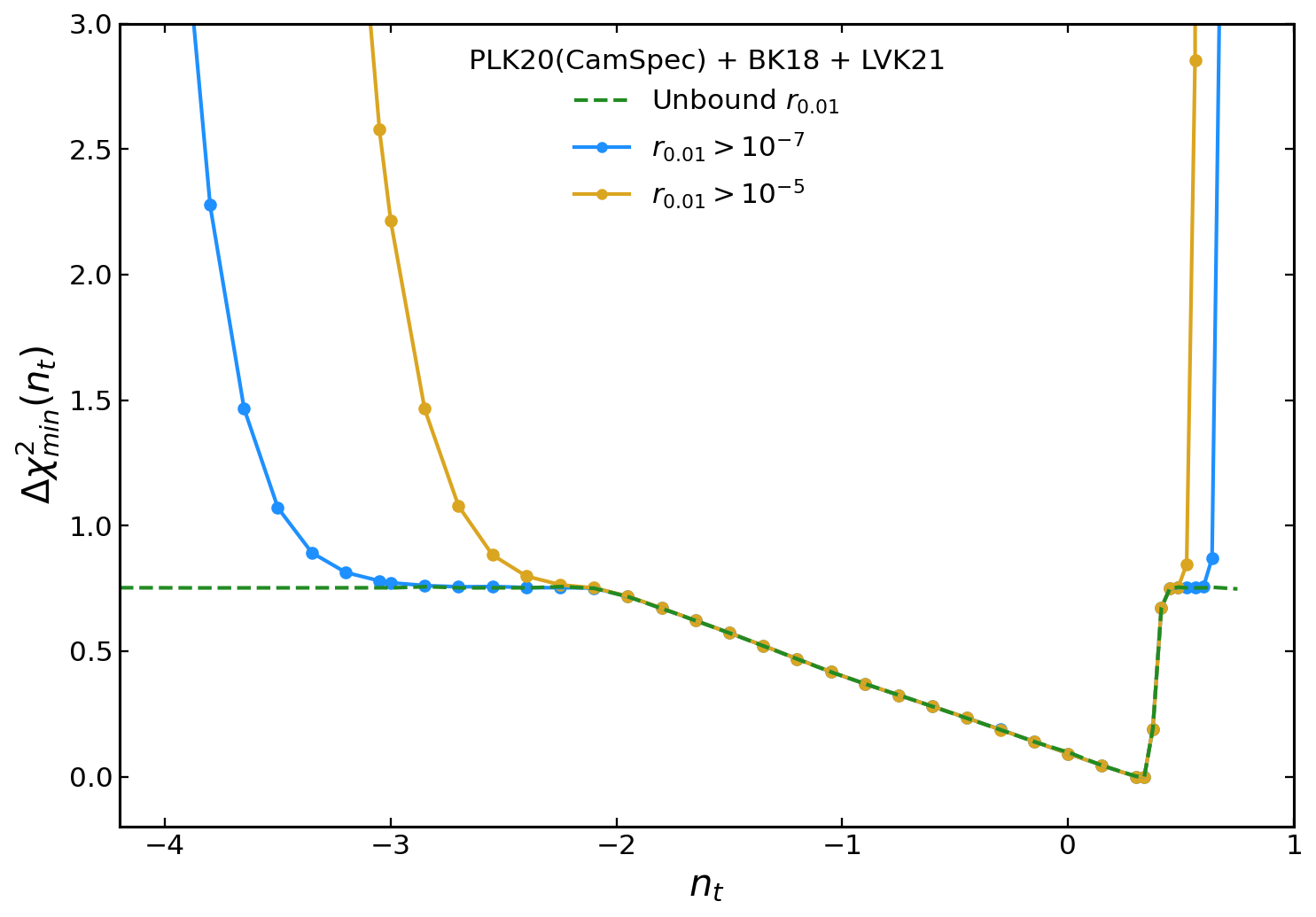}
    \hfill
    \includegraphics[width =0.49\textwidth]{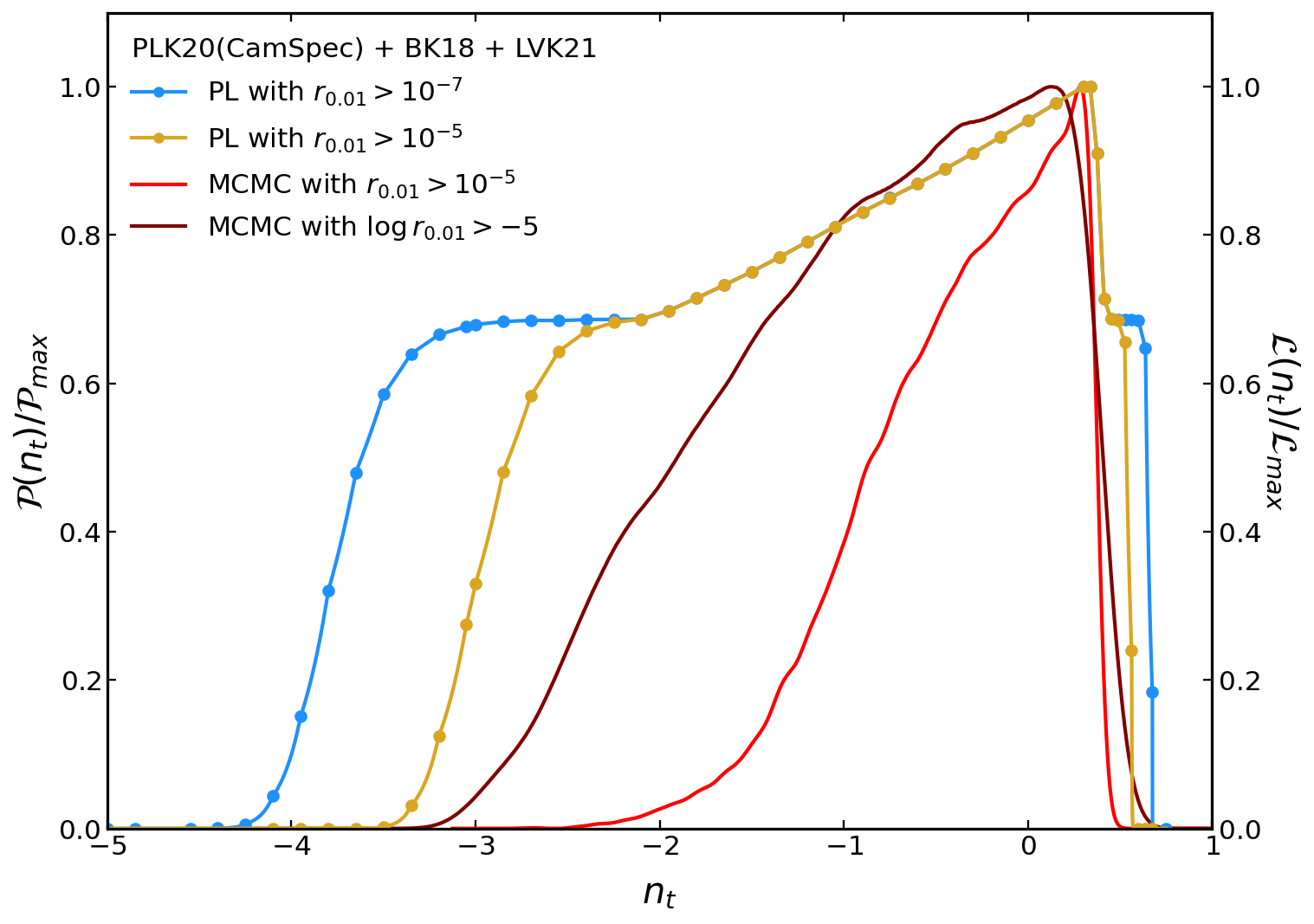}
    \caption{\textit{Left}: Comparison between $\Delta\chi^2_{min}$ as a function of $n_t$ for PLK20(CamSpec) + BK18 + LVK21, while imposing $r_{0.01} > 10^{-7}$ (blue) and $r_{0.01} > 10^{-5}$ (gold). Here we also show the corresponding result from \autoref{fig:nt_profiles} (dashed green) to emphasize the effect of bounding $r_{0.01}$. \textit{Right}: Comparison between $\mathcal{L}/\mathcal{L}_{\rm max}$ of the cases mentioned above and $\mathcal{P}/\mathcal{P}_{\rm max}$ from the MCMC (red) of \autoref{fig:mcmc_camspec_hlp}. Here, we plot the posterior obtained assuming a uniform prior on $\log r_{0.01} \in \qty[-5, 0]$.}
    \label{fig:nt_bounds}
\end{figure*}

The addition of LVK21 constrains the value of $n_t$ to be lower than $n_t \lesssim 0.4$ due to the huge level arm in the frequency domain. Given the low level of statistical significance, we would not interpret this as a sign for a non-constant spectral index. To emphasize this cutoff given by LVK21, we normalize the PLK20(CamSpec/HiLLiPoP) + BK18 + LVK21 profiles to the minimum of their counterpart without GW data. This also emphasizes that the corresponding plateaus are found at the same value of $\chi^2_{\rm min}$. Furthermore, LVK21 also causes another plateau at the same $\chi^2_{min}$ level w.r.t. the previously mentioned one, but this time for blue tilts. 

Finally, we note that HiLLiPoP shows a ``bump'' at $n_t \sim -1$ and that the plateaus are slightly higher compared to the minimum $\chi^2$ when considering HiLLiPoP rather than CamSpec, suggesting that HiLLiPoP is slightly more constraining $n_t$ compared to CamSpec.

The plateaus observed in the PL result from the fact that the combination of likelihoods is no longer sensitive to the value of $n_t$ in this region. This can be understood by looking at the co-profile $r_{0.01}(n_t)$ (\autoref{fig:nt_coprofiles}).
Indeed, in order to accommodate extreme values of $n_t$, the tensor-to-scalar ratio is suppressed to very small values. At some point, the likelihoods are no longer sensitive to whether tensors are there or not, and the $\chi^2$ just sits on the same value for each tilt.

This is exactly the reason behind the analysis of \cite{Galloni:2023}. Indeed, if we do not impose a lower cutoff in $r_{0.01}$, the datasets we consider are not sufficient to constrain $n_t$ given that one can always find a value of $r_{0.01}$ low enough to accommodate any tilt. 
As a consequence, we do not have a clear upper or lower bound on the $n_t$ profile (\autoref{fig:nt_profiles}). On the contrary, introducing a cutoff allows to get constraints on $n_t$ as illustrated in \autoref{fig:nt_bounds}. First, note that as soon as the tensor-to-scalar ratio is restricted, $\Delta\chi^2_{\rm min}$ diverges, indicating that we could recover both an upper and a lower bound. In fact, with $r_{0.01} = 10^{-7}$ (resp. $10^{-5}$), at some point the tilt will be so red that the corresponding B-modes should be observed by \textit{Planck} or BICEP/Keck array. Despite this, once again we stop at a qualitative description of the profile in $n_t$, as a further study is necessary for an actual confidence interval.

The same reason explains the difference between the posterior and the profile on $n_t$ (see \autoref{fig:nt_bounds}). Indeed, the values of $r_{0.01}$ found by the PL minimization are so low ($r_{0.01} \sim 10^{-10}-10^{-17}$) that it would be impossible to explore that region with our MCMC using a flat prior.
The region explored by the MCMC corresponds to the region of the co-profile with the highest values of $r_{0.01}$, which also corresponds to the region of the PL between the plateaus (see \autoref{fig:nt_bounds}).
Indeed, it is a well-known feature of uniform priors to poorly explore the region where the parameter of interest is extremely close to the boundaries. To showcase this, we show in the right panel of \autoref{fig:nt_bounds} the posterior obtained imposing a log-uniform prior on $r_{0.01}$. This allows us to better explore the region of very low tensor-to-scalar ratio and results in a broader distribution of $n_t$. 

\begin{figure*}[t]
    \centering
    \includegraphics[width = 0.49\textwidth]{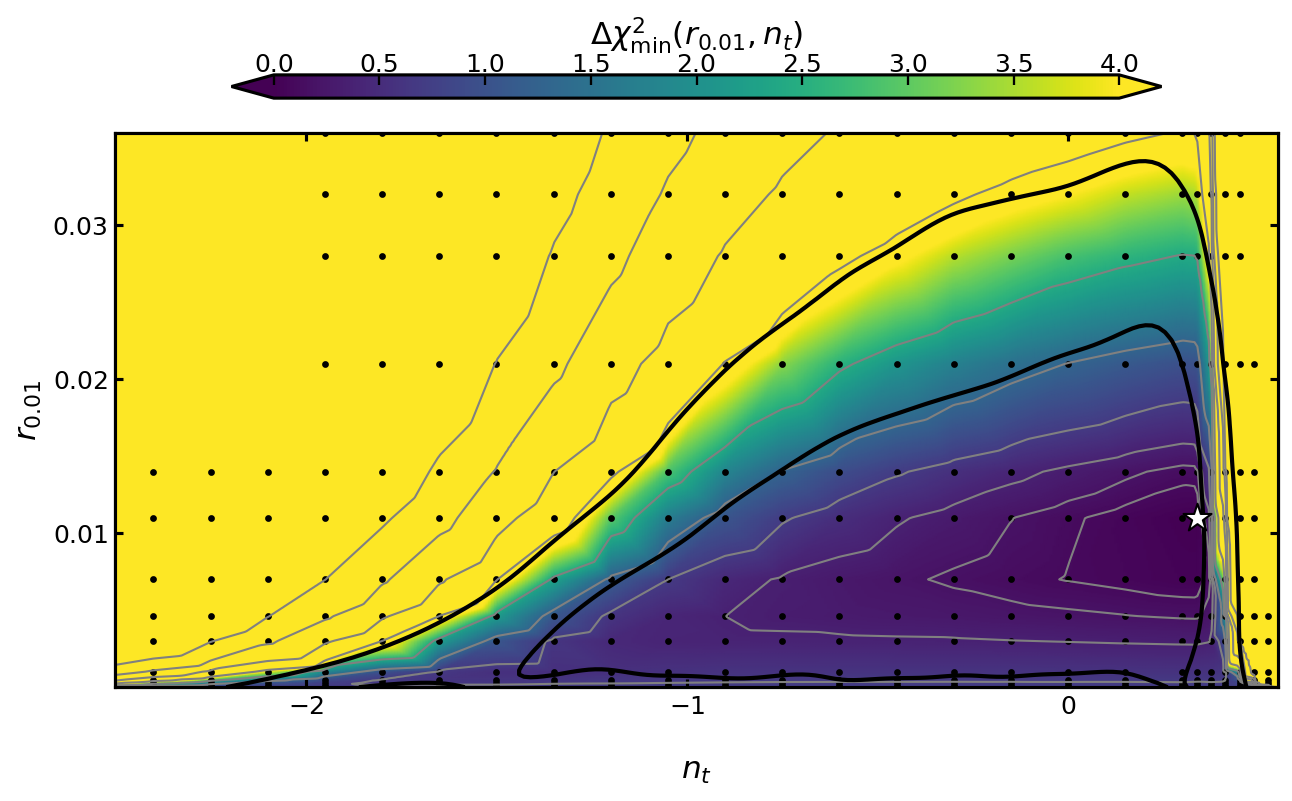}
    \hfill
    \includegraphics[width = 0.49\textwidth]{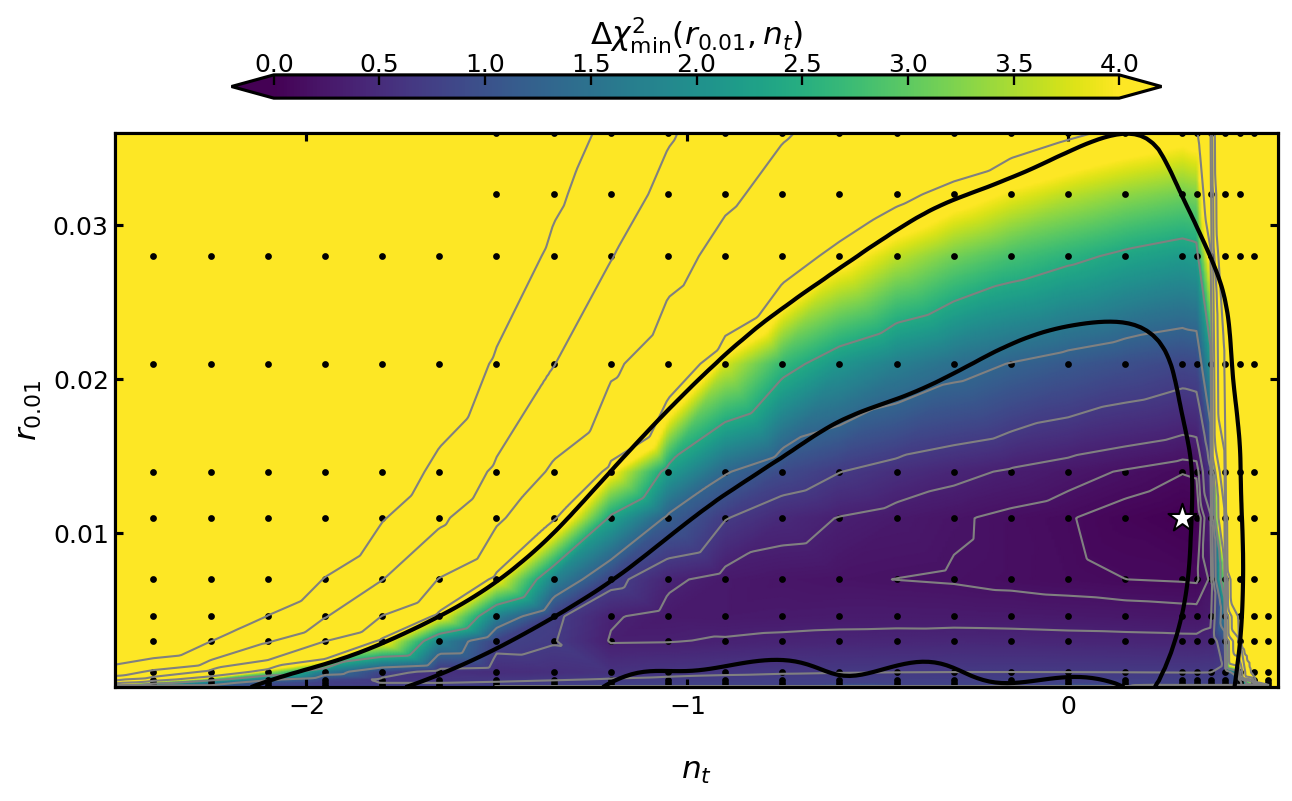}
    \caption{2D PL on $r_{0.01}$ and $n_t$ using PLK20(CamSpec) + BK18 + LVK21 (left) and PLK20(HiLLiPoP) + BK18 + LVK21 (right). The star indicates the absolute minimum we find with this procedure, whereas the gray lines show some iso-$\chi^2$ curves to emphasize the 2D shape of the profile. The black dots are the points in which a minimization is performed. We show in solid black the 68\% and 95\% contours of the 2D marginalized posterior from \autoref{sec:pl_mcmc_res}.}
    \label{fig:2d_profile}
\end{figure*}

Once again, this comparison highlights a possible flaw of the Bayesian analysis: a tilt of $n_t = -2$ is almost excluded by our MCMC analysis, however, it is not statistically excluded by the PL analysis as it only corresponds to $\Delta\chi^2_{min} \simeq 0.7$. This is not the case if we look at the log-uniform results. In fact, values as low as $n_t \simeq -3$ are still accepted and correspond to $\Delta\chi^2_{min} \gtrsim 3$, due to a better representation of the low-$r$ region. Despite this, such a choice is known to provide extremely underestimated bounds on $r_{0.01}$ \cite{Hergt:2021qlh}.

\subsection{2D profile likelihood}\label{sec:res_2D_PL}

In order to explore deeper the $(r_{0.01},n_t)$ plane, we construct the 2D profile likelihood, fitting the best fit over the rest of the parameters for any fixed combination of $r_{0.01}$ and $n_t$.

Of course, such an analysis is very demanding in terms of CPU-time, given that we have to iterate an already computationally heavy procedure. For this reason, we perform it only on PLK20(CamSpec/HiLLiPoP) + BK18 + LVK21 and not on their counterparts without LVK21.
We assume a range of the tensor-to-scalar ratio from a maximum of $r_{0.01} = 0.036$ to a minimum of $r_{0.01} = 10^{-6}$, for a total of $N = 15$ steps. The range in $n_t$ varies case by case, as \autoref{fig:2d_profile} shows (black dots). Then, we interpolate these points to get a smooth surface that represents the result for $N \to \infty$.

We overplot the marginalized 2D posteriors from the Bayesian analysis discussed in \autoref{sec:pl_mcmc_res}. The posteriors seem to follow the same behavior as the iso-$\chi^2$ curves, at least for $r_{0.01} \gtrsim 5 \times 10^{-4}$. In particular, both 95\% posterior contours follow approximately the iso-$\chi^2$ curve at $\Delta\chi^2_{\rm min} \simeq 4.6$. Similarly to the discussion at the end of \autoref{sec:res_PL_nt}, this depends on the choice of the prior for the MCMC exploration, thus this correspondence is not guaranteed if the SSA is not employed.

Here, if we imagine intersecting this surface with any horizontal (vertical) plane for a value of $r_{0.01}$ ($n_t$), we would obtain the PL of $n_t$ ($r_{0.01}$) conditioned on that value.
We see that for the values of $r_{0.01}$ considered here, the PL do not show any plateau (see \autoref{fig:nt_profiles}), and instead $\chi^2$ diverges for low and high values of $n_t$ allowing in principle to derive some confidence interval.

\begin{figure*}[t]
    \centering
    \includegraphics[width = 0.49\textwidth]{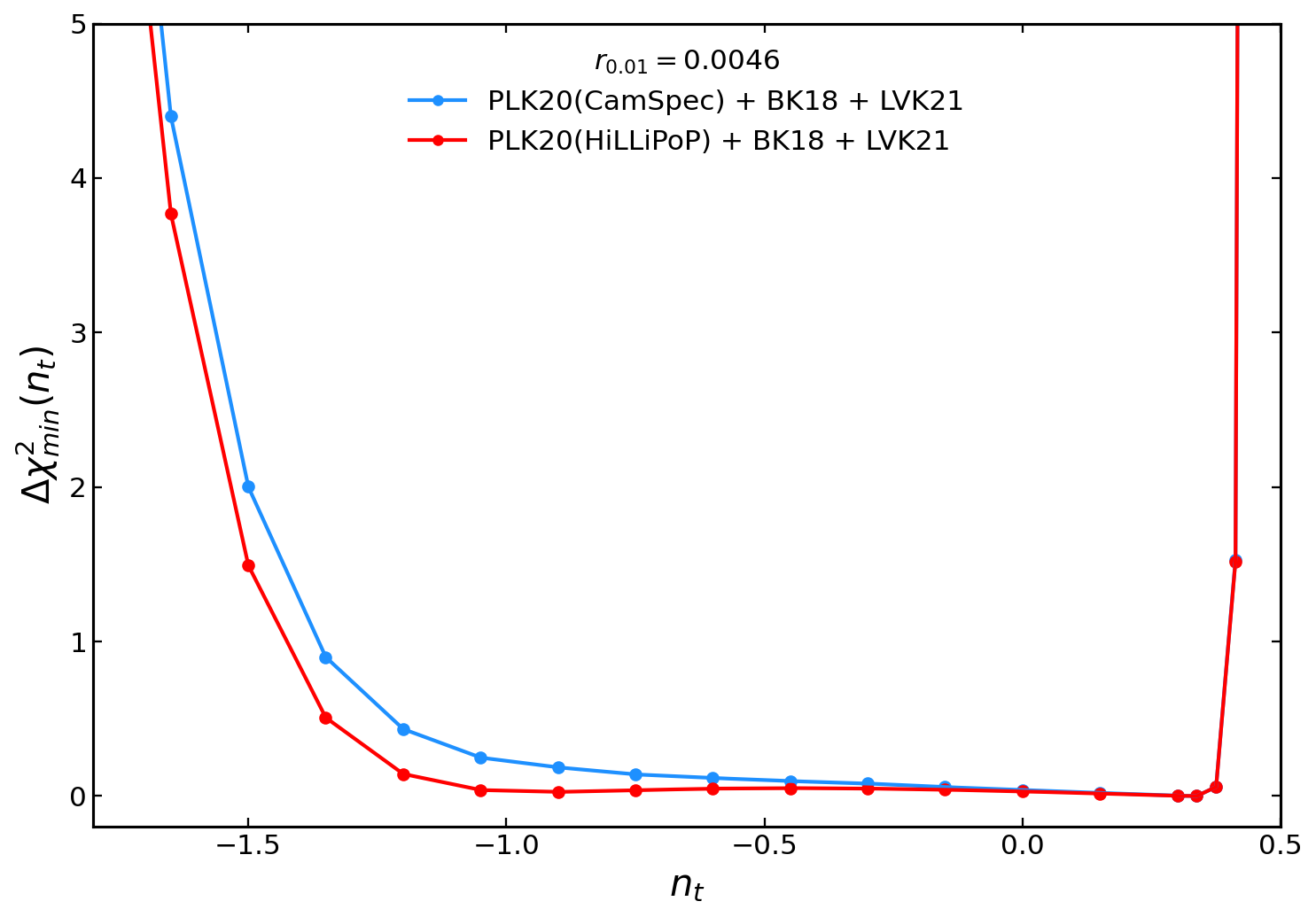}
    \hfill
    \includegraphics[width = 0.49\textwidth]{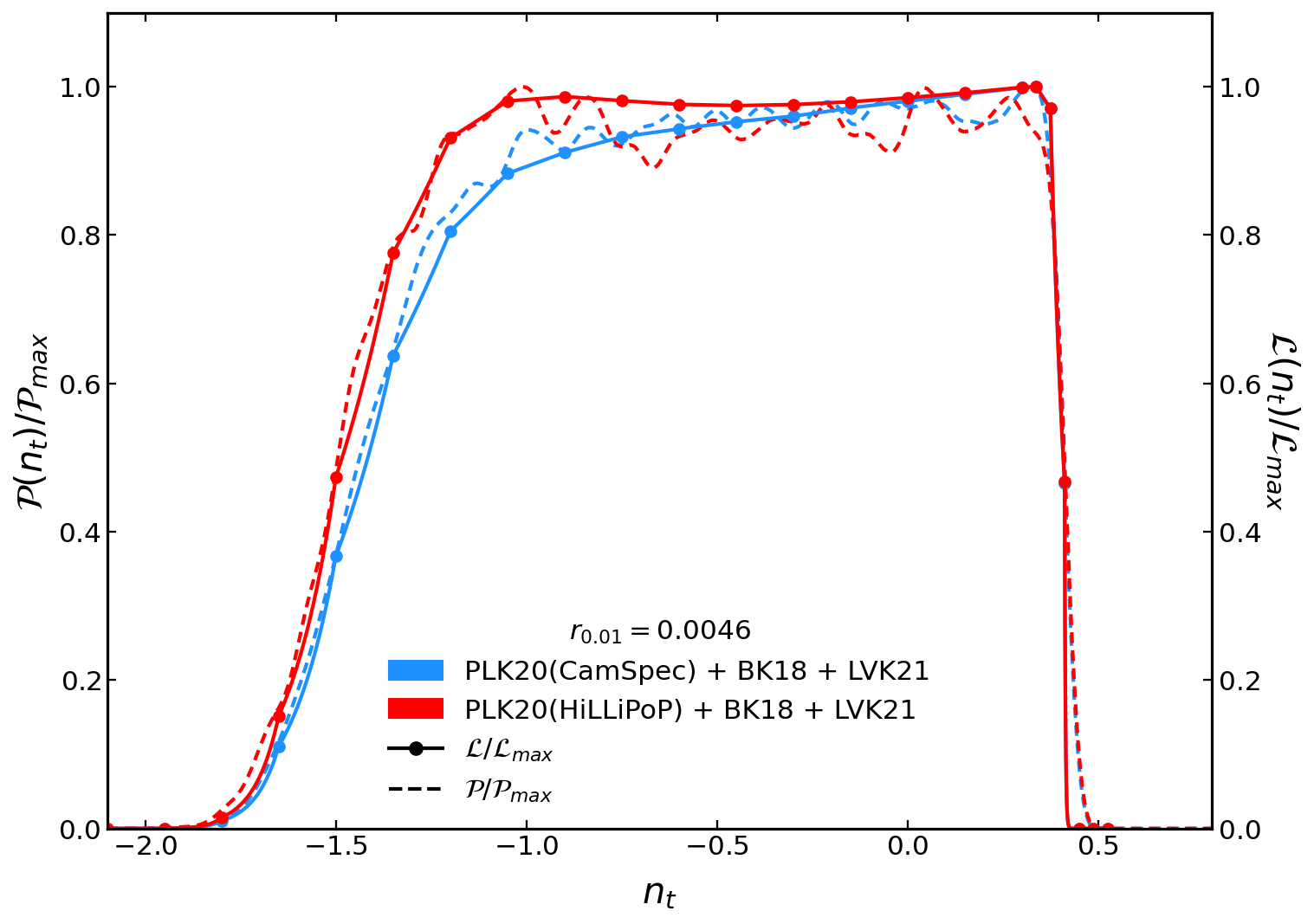}
    \caption{$\Delta\chi^2_{min}$ as a function of $n_t$, while imposing $r_{0.01} = 0.0046$ corresponding to Starobinsky inflation. In red we plot the results of PLK20(CamSpec) + BK18 + LVK21, while in blue the ones of PLK20(HiLLiPoP) + BK18 + LVK21.}
    \label{fig:nt_bounds_staro}
\end{figure*}

To illustrate the results, we fix the tensor-to-scalar ratio to the value predicted by the Starobinsky inflation $r_{0.01} \sim 0.0046$ \cite{Starobinsky}. The resulting profiles on $n_t$ are shown in \autoref{fig:nt_bounds_staro} for PLK20(HiLLiPoP) + BK18 + LVK21 (solid blue) and PLK20(CamSpec) + BK18 + LVK21 (solid red). On top of this, we obtain the Bayesian equivalent of this by running an MCMC imposing again $r_{0.01} = 0.0046$. The resulting posterior distributions are shown in the same figure with dashed lines. 

Note that the frequentist and Bayesian results agree quite well in this case. As in the full 2D case, this correspondence indicates that the other parameters do not induce any volume effects on $n_t$. This also ensures that the credible and confidence intervals should be the same, allowing us to obtain frequentist confidence intervals. 

Of course, we do not know a priori the $\Delta\chi_{\rm min}^2$ corresponding to the 95\% CL, since the distribution is clearly non-Gaussian. Still, we can obtain it from the Bayesian intervals by finding the $\Delta\chi_{\rm min}^2$ that gives the same values. In particular, the Bayesian intervals read
\begin{equation*}
\begin{aligned}
    -1.43 < n_t < 0.41 &\qq{PLK20(CamSpec)+BK18+LVK21},\\
    -1.47 < n_t < 0.41 &\qq{PLK20(Hillipop)+BK18+LVK21},
\end{aligned}
\end{equation*}
which correspond respectively to $\Delta\chi_{\rm min}^2 = 1.5$ and $\Delta\chi_{\rm min}^2 = 1.3$. As expected, these are lower than the corresponding Gaussian prescription for the 95\% CL interval of $\Delta\chi_{\rm min}^2 = 3.84$, since both of our distributions are much flatter than a Gaussian.

As a final remark, HiLLiPoP has a mild preference for $n_t \simeq -1$, also shown in \autoref{fig:nt_profiles}, which is not shown by CamSpec.

\section{Conclusions} \label{sec: conclusions}

\begin{figure}
    \centering
    \includegraphics[width=\hsize]{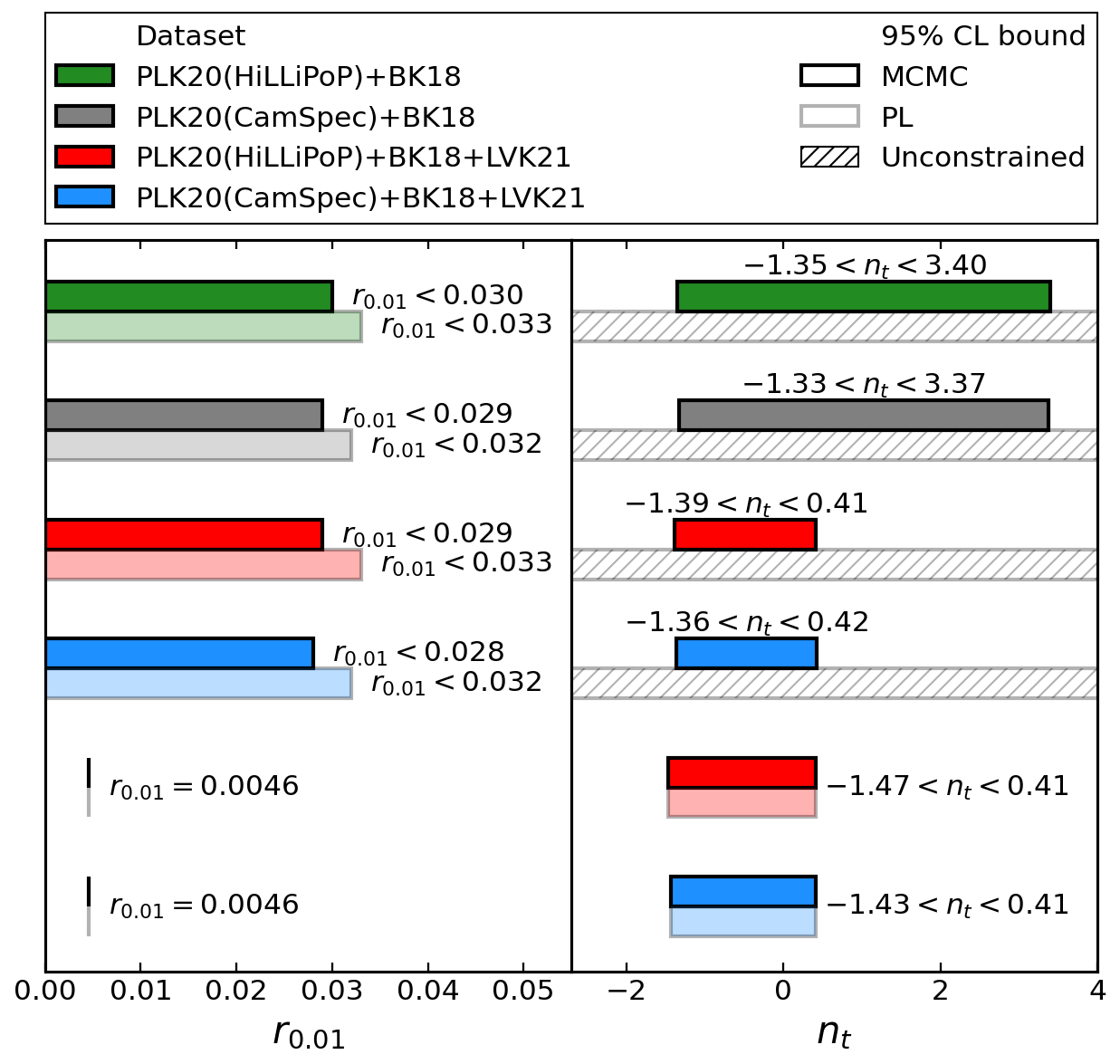}
    \caption{Summary of the 95\% CL intervals obtained in this work. The solid patches refer to the credible intervals, while the shaded ones to the confidence intervals. The last two rows refer to $r_{0.01} = 0.0046$ as predicted by Starobinsky inflation.}
    \label{fig:summary}
\end{figure}

In this work, we explore the primordial tensor perturbations constraints utilizing the latest available data from CMB and Gravitational Waves measurements. We focus more specifically on the tensor-to-scalar ratio $r_{0.01}$ and the tensor spectral tilt $n_t$ parametrizing the amplitude and the tilt of the tensor perturbations power spectrum. Our dataset includes CMB measurements from the latest \textit{Planck} PR4 dataset, the BICEP/Keck array and LIGO-Virgo-KAGRA upper bound on the energy density of GWs to constraint the tensor spectrum at much smaller scales.

In order to provide a complete picture of the constraints, we use both Bayesian and frequentist approaches. While the former provides with a statistical statement about the probability distribution of the different parameters after marginalization, the latter can describe the probability distribution of data given the theoretical model insensitive to the choice of priors and free from volume effects. Thus, it is clear that comparing credible and confidence intervals is not trivial; however, it allows us to extract useful information on relatively unconstrained parameters. The results are summarized in \autoref{fig:summary}.

For the tensor-to-scalar ratio, we find upper-limits with Profile Likelihoods slightly more conservative than the Bayesian ones. This indicates that both volume-effects and prior choices have a role in obtaining the credible intervals. 

Concerning the tensor spectral tilt $n_t$, the recovered profile likelihoods exhibit a very non-Gaussian behavior with the absence of distinct bounds. In fact, $\Delta\chi^2_{\rm min}$ never exceeds $\Delta\chi^2_{\rm min} \simeq 1$, which cannot be captured by Bayesian analysis.
We identify that this behavior is the result of the potential accommodation of highly extreme spectral tilts at the expense of driving $r_{0.01}$ toward zero. Despite this, datasets accounting for GW interferometers show a mild preference for $n_t \simeq 0.3$. Instead, those without LIGO-Virgo-KAGRA prefer $n_t \simeq 2$.
Some constraints on $n_t$ can be found by introducing a cutoff on small values of $r_{0.01}$ as done in Bayesian analysis ($r_{0.01} > 10^{-5}$).
However, it cannot solve the discrepancy with the marginalized posterior of the MCMC analysis when it is driven by a flat prior on $r_{0.01}$. For example, sampling with a log-uniform prior allows us to explore the region for very low values of $r_{0.01}$ and provide a marginalized posterior significantly wider. Indeed, in the case of $n_t$, given today's accuracy of the data, Bayesian analysis is dominated by the choice of priors and consequently credibility intervals are driven by this choice. In contrast, frequentist analysis, which is independent of any prior, leads to unconstrained $n_t$.

When profiling $r_{0.01}$ and $n_t$ simultaneously in 2D, we recover constraints close to the 2D marginalized posterior for large values of $r$ ($r_{0.01}\gtrsim5\times10^{-4}$). This indicates that other parameters (physical or nuisance) do not induce volume effects in this 2D plane, proving that in this range the tensor sector is sufficiently decoupled from the $\Lambda$CDM parameters. For the largest value of $r_{0.01}$, including the prediction for the Starobinsky model ($r_{0.01}=0.0046$), constraints on $n_t$ can be derived (see \autoref{fig:summary}). In this case, the correspondence between Bayesian and frequentist results is evident, ensuring that credible and confidence intervals are the same.

Projecting this comparison into the future, when more data will be available, we can expect similar difference between these two methods as they answer different questions, especially in the case of an upper-limit on $r$. Indeed, the prior dependence of the Bayesian analysis is an unavoidable part of the framework, so one must deal with the fact that a prior must be taken before performing any analysis. Instead, performing the same analysis with different priors is a way to showcase the consequence of such a choice \cite{Galloni:2023}. On the other hand, the more the new data will be constraining on the tensor sector, the more frequentist and Bayesian results will converge to each other. This is because the likelihood function will dominate both methods in the limit of a full detection of $r$.

To summarize, this paper delivers a comprehensive and statistically robust analysis of the tensor sector of parameter space. Beyond offering an updated perspective using the latest datasets currently available, the analysis underscores the significance of probing relatively unconstrained parameters with frequentist approaches, complementing the widely used Bayesian methods. This insight proves valuable for forthcoming investigations into CMB polarization, such as with LiteBIRD, and for any endeavor exploring extensions to the standard $\Lambda$CDM model.

\acknowledgments

The authors thank Gilles Weymann-Despres and St\'ephane Ili\'c for useful comments and discussions. This work is based on observations obtained with \textit{Planck} (\url{http://www.esa.int/Planck}), an ESA science mission with instruments and contributions directly funded by ESA Member States, NASA, and Canada. 
The \textit{Planck} PR4 data are publicly available on the \textit{Planck} Legacy Archive (\url{http://pla.esac.esa.int}). Both likelihoods LoLLiPoP and HiLLiPoP based on PR4 are publicly available on github (\url{http://github.com/planck-npipe}) as external likelihoods for \texttt{Cobaya}.
Some of the results in this paper have been derived using the following packages: \texttt{CAMB} \cite{Lewis:2013hha, Lewis_2000, Howlett_2012}, \texttt{Cobaya} \cite{Torrado:2020xyz}, \texttt{GetDist} \cite{Lewis:2019}, \texttt{Matplotlib}\footnote{\url{https://github.com/matplotlib/matplotlib}.} \cite{matplotlib} and \texttt{NumPy}\footnote{\url{https://github.com/numpy/numpy}.} \cite{numpy}. G.G. acknowledges support from ASI/LiteBIRD grant number 2020-9-HH.0, and by the InDark and LiteBIRD INFN projects.

\appendix

\section{Constraints on the \texorpdfstring{$\Lambda$}{TEXT}CDM parameters} \label{app:LCDM}

As mentioned in \autoref{sec: methodology}, the focus of this work is the tensor sector of parameter space, which consists of $r_{0.01}$ and $n_t$. Despite this, to correctly capture the variability in these parameters, it is also important to sample all other $\Lambda$CDM and nuisance parameters. First, we define a set of priors in \autoref{tab:priors}, encoding our knowledge about physical parameters. For what regards the nuisance parameters of the various likelihoods, we stick to the default settings.

\begin{table}[t]
\caption{\label{tab:priors}
Priors on the parameters of $\Lambda$CDM + $r_{0.01}$ + $n_t$. Here, $A_{\rm s}$ is the scalar perturbations amplitude, $n_{\rm s}$ the scalar spectral tilt, $\Omega_{\rm b}$ and $\Omega_{\rm cdm}$ are the abundances of baryons and CDM, $h \equiv H_0/100$ is the Hubble constant divided by 100, $\tau_{\rm reio}$ the optical depth and $\theta_{\rm MC}$ is an approximate quantity representing the sound horizon.}
\begin{ruledtabular}
\begin{tabular}{lccc}
\textrm{Parameter}& \textrm{Prior} & \textrm{Parameter}& \textrm{Prior}\\
\colrule
$\log(10^{10}A_{\rm s})$ & ${[}1.61, 3.91{]}$  & $\tau_{\rm reio}$ & ${[}0.01, 0.8{]}$ \\
$n_{\rm s}$             & ${[}0.8, 1.2{]}$    & $\theta_{\rm MC}$ & ${[}0.5, 10{]}$ \\
$\Omega_{\rm b} h^2$    & ${[}0.005, 0.1{]}$  & $r_{0.01}$        & ${[}10^{-5}, 3{]}$ \\
$\Omega_{\rm cdm} h^2$  & ${[}0.001, 0.99{]}$ & $n_{\rm t}$       & ${[}-5, 5{]}$ \\
\end{tabular}
\end{ruledtabular}
\end{table}
In addition, we fix the number of relativistic degrees of freedom to $N_{\rm eff} = 3.046$, asking for a massive neutrino with $M_\nu = 0.06$ eV.

\begin{figure}[t]
    \centering
    \includegraphics[width = \hsize]{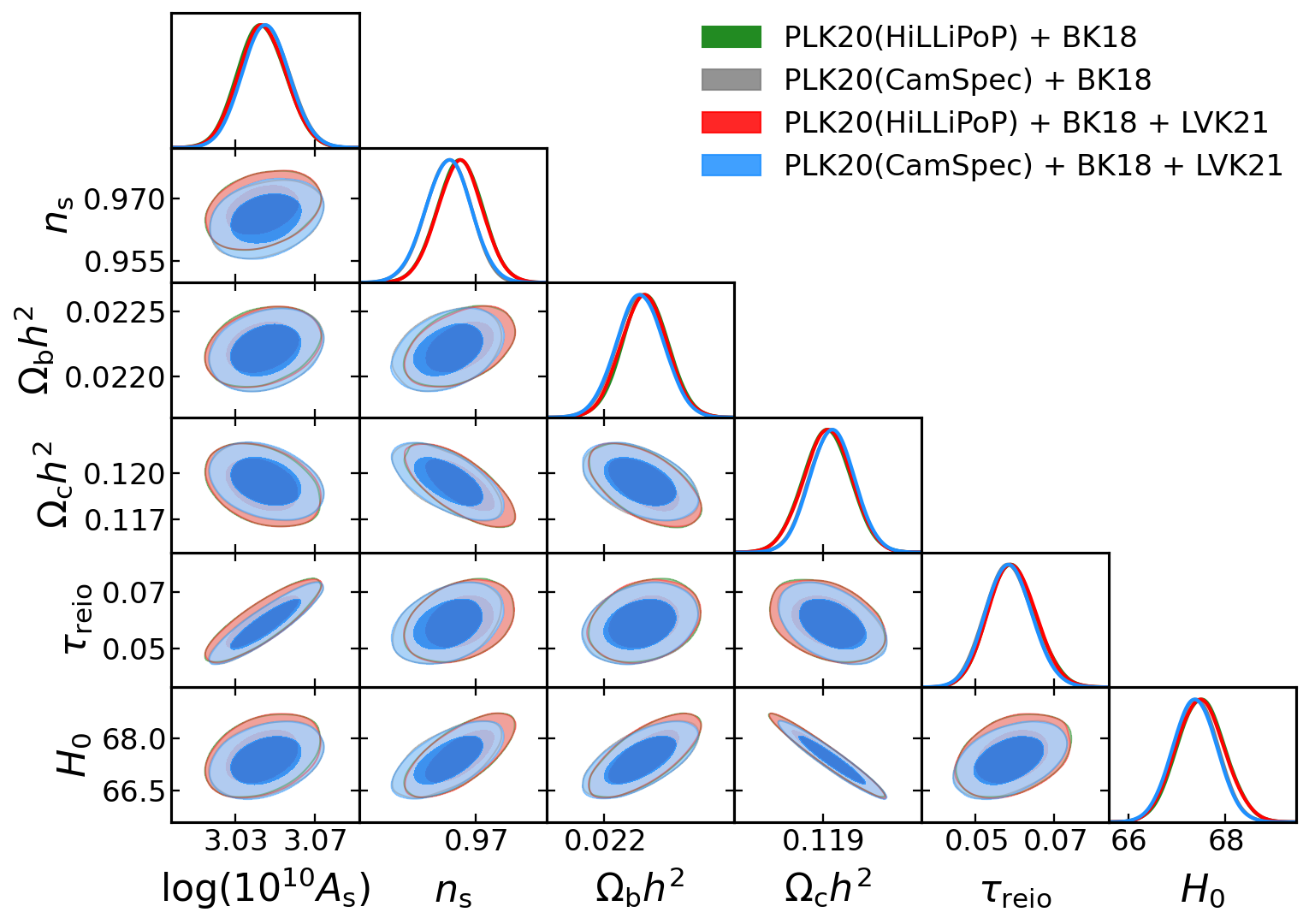}
    \caption{$\Lambda$CDM contours}
    \label{fig:lcdm_triangle}
\end{figure}

In \autoref{sec:pl_mcmc_res}, the results on the tensor sectors are discussed, while here we show those on the $\Lambda$CDM parameters. In fact, \autoref{fig:lcdm_triangle} shows the triangle plot of what we may call the ``scalar'' sector of parameters. Note that here we substitute $\theta_{\rm MC}$ with $H_{\rm 0}$ as it is a more physical parameter controlling the sound horizon. 
Here, we consider PLK(CamSpec) + BK18 + LVK21, PLK(CamSpec) + BK18, PLK(HiLLiPoP) + BK18 + LVK21, and PLK(HiLLiPoP) + BK18 showing that they are all compatible with each other. 

To be more quantitative, we also report in \autoref{tab:LCDM_constraints} the mean values and standard deviations of these parameters.
\begin{table*}[t]
\caption{\label{tab:LCDM_constraints}
Summary of the constraints on the $\Lambda$CDM + $r_{\rm 0.01}$ + $n_{\rm t}$ model. For the 6 $\Lambda$CDM parameters we report the mean and the standard deviation; instead, we report the 95\% CL intervals for the tensor sector, as in \autoref{sec:pl_mcmc_res}.}
\footnotesize
\begin{ruledtabular}
\begin{tabular}{lcccc}
\textrm{Parameter}& \textrm{PLK20(CamSpec)+BK18+LVK21} & \textrm{PLK20(HiLLiPoP)+BK18+LVK21}& \textrm{PLK20(CamSpec)+BK18} & \textrm{PLK20(HiLLiPoP)+BK18}\\
\colrule
$\log(10^{10}A_{\rm s})$ & $3.046 \pm 0.012$     & $3.044 \pm 0.012$     & $3.045 \pm 0.012$ & $3.044 \pm 0.012$ \\
$n_{\rm s}$             & $0.9653 \pm 0.0039$   & $0.9673 \pm 0.0039$   & $0.9653 \pm 0.0038$ & $0.9673 \pm 0.0039$ \\
$\Omega_{\rm b} h^2$    & $0.02220 \pm 0.00013$ & $0.02222 \pm 0.00013$ & $0.02220 \pm 0.00013$ & $0.00222 \pm 0.00013$\\
$\Omega_{\rm cdm} h^2$  & $0.1194 \pm 0.0010$   & $0.1192 \pm 0.0011$   & $0.1194 \pm 0.0010$ & $0.1192 \pm 0.0011$\\
$\tau_{\rm reio}$       & $0.0586 \pm 0.0059$   & $0.0594 \pm 0.0060$   & $0.0586 \pm 0.0059$ & $0.0594 \pm 0.0060$\\
$H_{\rm 0}$             & $67.37 \pm 0.46$      & $67.49 \pm 0.50$      & $67.40 \pm 0.45$ & $67.50 \pm 0.49$  \\ 
$r_{\rm 0.01}$       & $< 0.028$   & $< 0.029$   & $< 0.029$ & $< 0.030$\\
$n_{\rm t}$             & $-1.36 < n_{\rm t} < 0.42$      & $-1.39 < n_{\rm t} < 0.41$      & $-1.33 < n_{\rm t} < 3.37$ & $-1.35 < n_{\rm t} < 3.40$  \\
\end{tabular}
\end{ruledtabular}
\end{table*}

As expected, removing LVK21 does not produce any difference in $\Lambda$CDM parameters. The only noticeable difference is found by looking at $n_s$ for PLK20(CamSpec) and PLK20(HiLLiPoP), which is estimated to be slightly higher (approximately $0.5\sigma_{n_s}$ higher) by the latter. This feature is also remarked in \cite{Tristram:2023}. For completeness, we also mention that HiLLiPoP also estimates $H_0$ to be higher of $\sim 0.2\sigma_{H_0}$.

\section{Tests on the Accuracy} \label{app:accuracy_pl}

In \autoref{sec: methodology} we mention that to perform a precise PL, it is crucial to efficiently maximize the likelihood, reaching at each point of the profiled parameter the absolute minimum of the $\chi^2$ (and not a local one). Thus, to verify the precision of our minimizing procedure, we performed different tests. For example, we fix the tensor-to-scalar ratio to $r=0.02$ and minimize the $\chi^2$ eight times, storing the result of each. We repeat this while changing the accuracy parameters of \texttt{CAMB} and of \texttt{MINUIT}. The dispersion of the results is an indication of the precision of minimization performed by \texttt{MINUIT}, since it gauges the reliability of recovering the absolute minimum. In addition, the value at which these points tend to depend to some degree on the accuracy of the Boltzmann solver \cite{planck2020-LVII}.

In particular, the parameters we considered for these tests are
\begin{itemize}
    \item \texttt{stra} of \texttt{MINUIT}, which allows to change the \say{strategy} of the minimization from \textit{fast} (\texttt{stra} $= 0$), \textit{balanced} (\texttt{stra} $= 1$) and \textit{accurate} (\texttt{stra} $= 2$). The default value is \texttt{stra} $= 1$.
    \item \texttt{AccuracyBoost} of \texttt{CAMB}, which controls several other accuracy parameters of the Boltzmann solver. The default value is $1$.
    \item \texttt{lAccuracyBoost} of \texttt{CAMB}, which is related to the resolution in $\ell$ space of the Boltzmann solver. The default value is $1$.
\end{itemize}
\begin{figure}[t]
    \centering
    \includegraphics[width = \hsize]{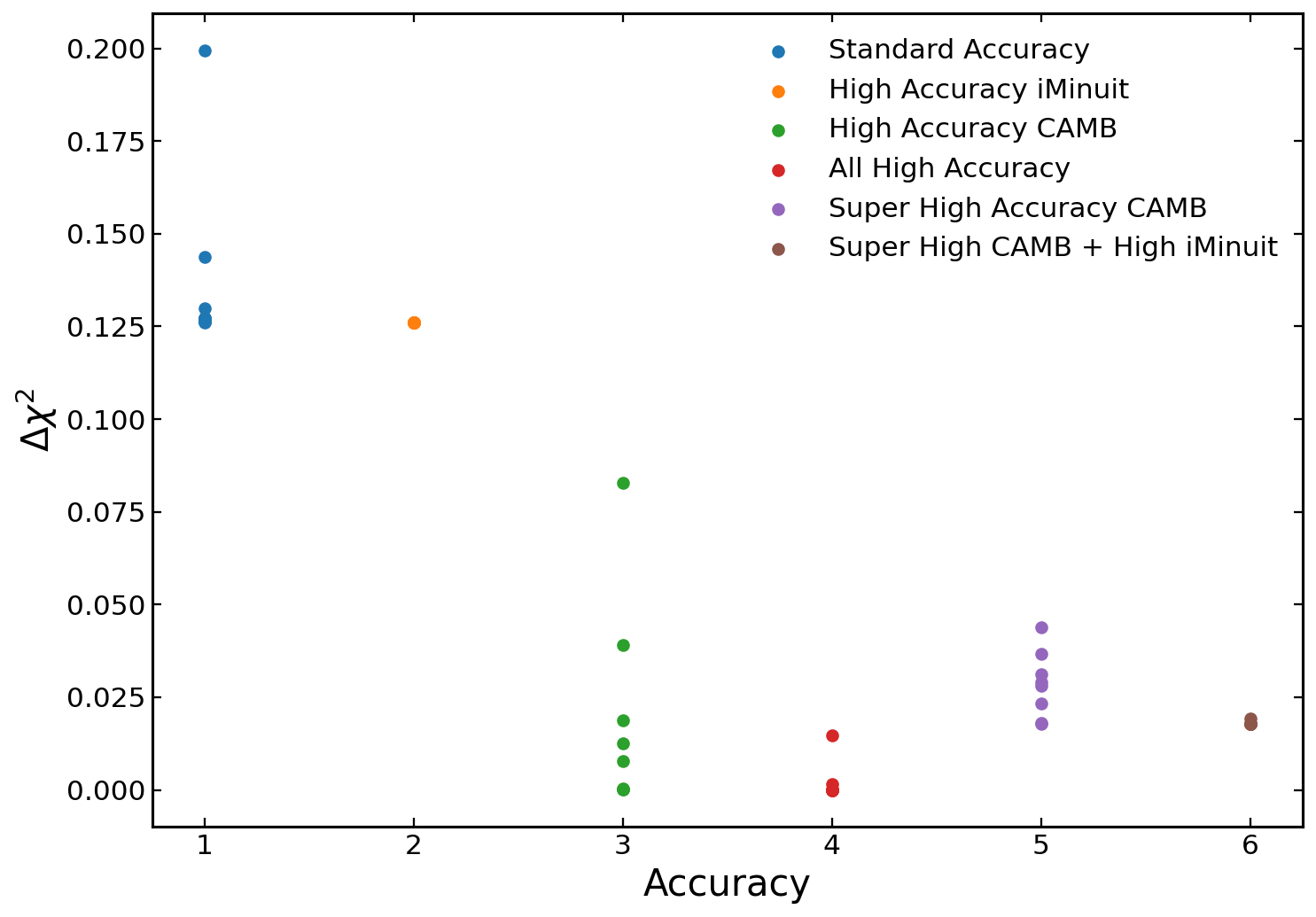}
    \caption{Different set of minima obtained with different settings for the accuracy of \texttt{CAMB} and \texttt{MINUIT}.}
    \label{fig:accuracies}
\end{figure}
\autoref{fig:accuracies} shows the different sets of minima we obtain, where we define
\begin{description}
    \item[Standard Accuracy] Default settings;
    \item[High Accuracy iMinuit] \texttt{stra} $=2$;
    \item[High Accuracy CAMB] \texttt{AccuracyBoost} $=2$ and \texttt{lAccuracyBoost} $=2$;
    \item[All High Accuracy] \texttt{stra} $=2$, \texttt{AccuracyBoost} $=2$ and \texttt{lAccuracyBoost} $=2$;
    \item[Super High Accuracy CAMB] \texttt{AccuracyBoost} $=3$ and \texttt{lAccuracyBoost} $=3$;
    \item[Super High CAMB + High iMinuit] \texttt{stra} $=2$, \texttt{AccuracyBoost} $=3$ and \texttt{lAccuracyBoost} $=3$.
\end{description}

It is clear that these settings have an effect on the values we obtain for the minimum. \autoref{fig:accuracies} shows that when we increase the accuracy of \texttt{MINUIT} the dispersion of the points gets severely reduced. Instead, the absolute minimum we obtain here is given by increasing the accuracy of \texttt{CAMB} to \textit{high}, since going to \textit{super high} actually increases the values of $\chi^2$. Despite this, all of these non-default settings increase significantly the computation time necessary to maximize the likelihood, thus we must ask ourselves if it is worth imposing some extra accuracy.

First, note that with \textit{high} \texttt{CAMB} accuracy, the typical $\chi^2$ decreases of $\sim 0.125$. Still, if this difference is just a constant offset between the two configurations, it will essentially disappear when we normalize the likelihood values on the absolute minimum. Furthermore, note that although increasing the accuracy of \textit{MINUIT} leads all points to converge to the absolute minimum, it is not worth it in terms of computation time. In fact, usually more than half of the points with the default \texttt{MINUIT} converge to the same point, so statistically we always get at least one point there.

Despite these considerations, we build a parabolic fit for each accuracy configuration (with $\order{10}$ points) to verify that we can stick to the default settings. We did not find significant differences in the final PL.

\section{Non-Gaussian Feldman-Cousins} \label{sec: generalized_fc}

In \autoref{sec:res_PL_r} we discuss the fact that our PLs on the tensor-to-scalar ratio present some non-Gaussian features. Thus, we ask ourselves whether it is possible to generalize the FC prescription to a more complex curve that fits our PLs.

We consider two simple alternatives that have two characteristics: we require that one of the parameters of the function corresponds to our parameter of interest; in addition, we will consider only this parameter as free to vary in the FC procedure. In other words, the curve we choose will shift to the right or to the left with the parameter changing, while keeping its shape unchanged. 

In particular, we consider what we call \textit{double Gaussian}, defined as
\begin{equation}
\begin{aligned}
    &P(x|\mu, \Delta\mu, \sigma_1, \sigma_2, A) \propto \\
    &\propto\exp\qty[-\frac{(x-\mu)^2}{2\sigma_1^2}] + A\exp\qty[-\frac{(x-\mu-\Delta\mu)^2}{2\sigma_2^2}]\ ,
    \label{eq:double_gaussian}
\end{aligned}
\end{equation}
and \textit{piecewise Gaussian}, which reads
\begin{equation}
\begin{aligned}
    &P(x|\mu, \Delta\mu, \sigma_1, \sigma_2, x_{sep}) \propto \\
    &\propto
    \begin{cases}
        \exp\qty[-\frac{(x-\mu)^2}{2\sigma_1^2}] \qq{for} x \leq x_{sep} \\
        \exp\qty[-\frac{(x-\mu-\Delta\mu)^2}{2\sigma_2^2}] \qq{for} x > x_{sep}
    \end{cases}\ .
    \label{eq:piecewise_gaussian}
\end{aligned}
\end{equation}

Of course, these curves do not find any physical justification, but are born from the pragmatic attempt to fit our PL with analytical formulas. The parameter values are estimated by fitting \autoref{eq:double_gaussian} and \autoref{eq:piecewise_gaussian} to our PL. These curves, together with the standard Gaussian, allow us to fit the features mentioned in \autoref{sec:res_PL_r}. 

\begin{figure*}
    \centering
    \includegraphics[width = 0.49\textwidth]{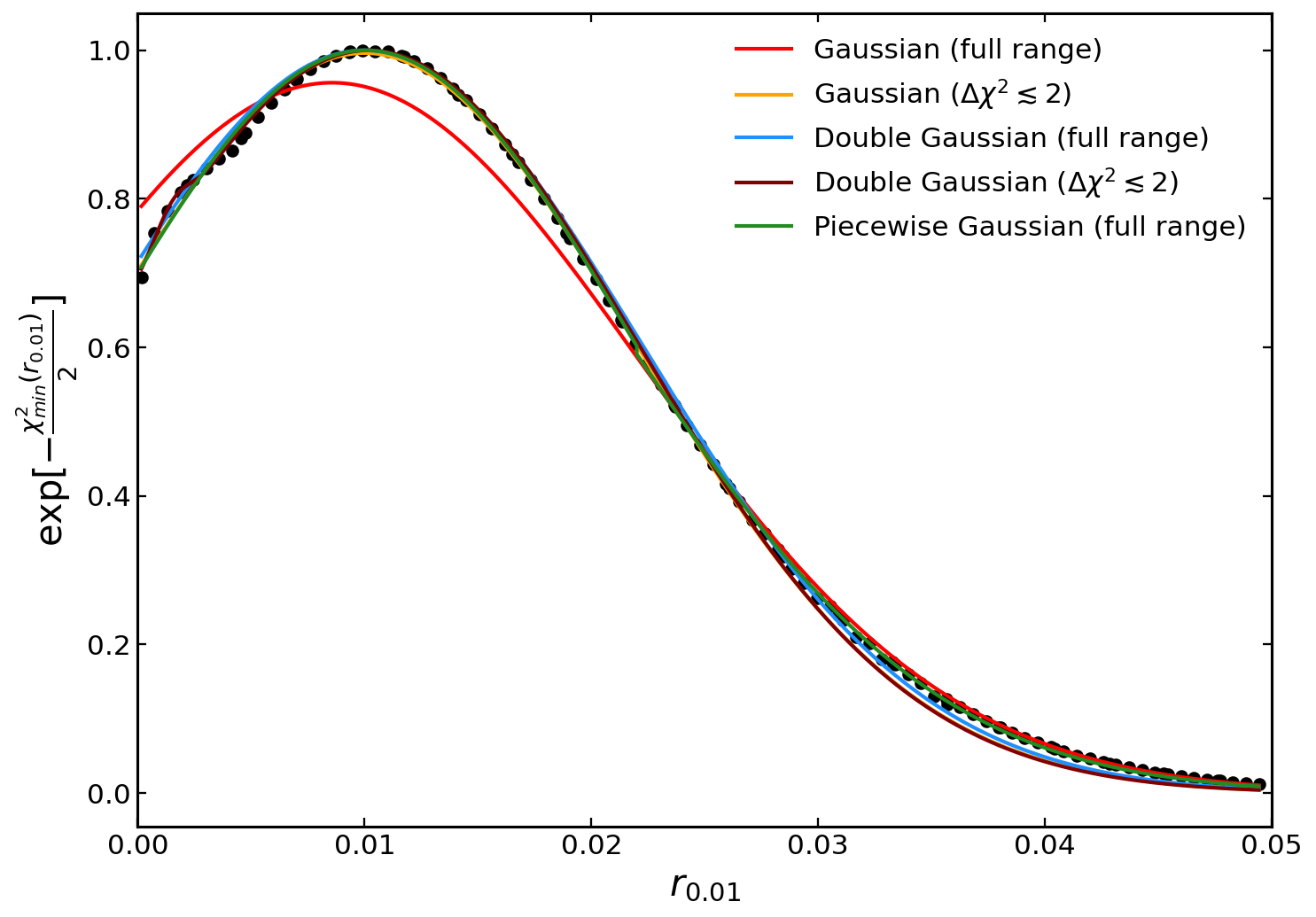}
    \hfill
    \includegraphics[width = 0.49\textwidth]{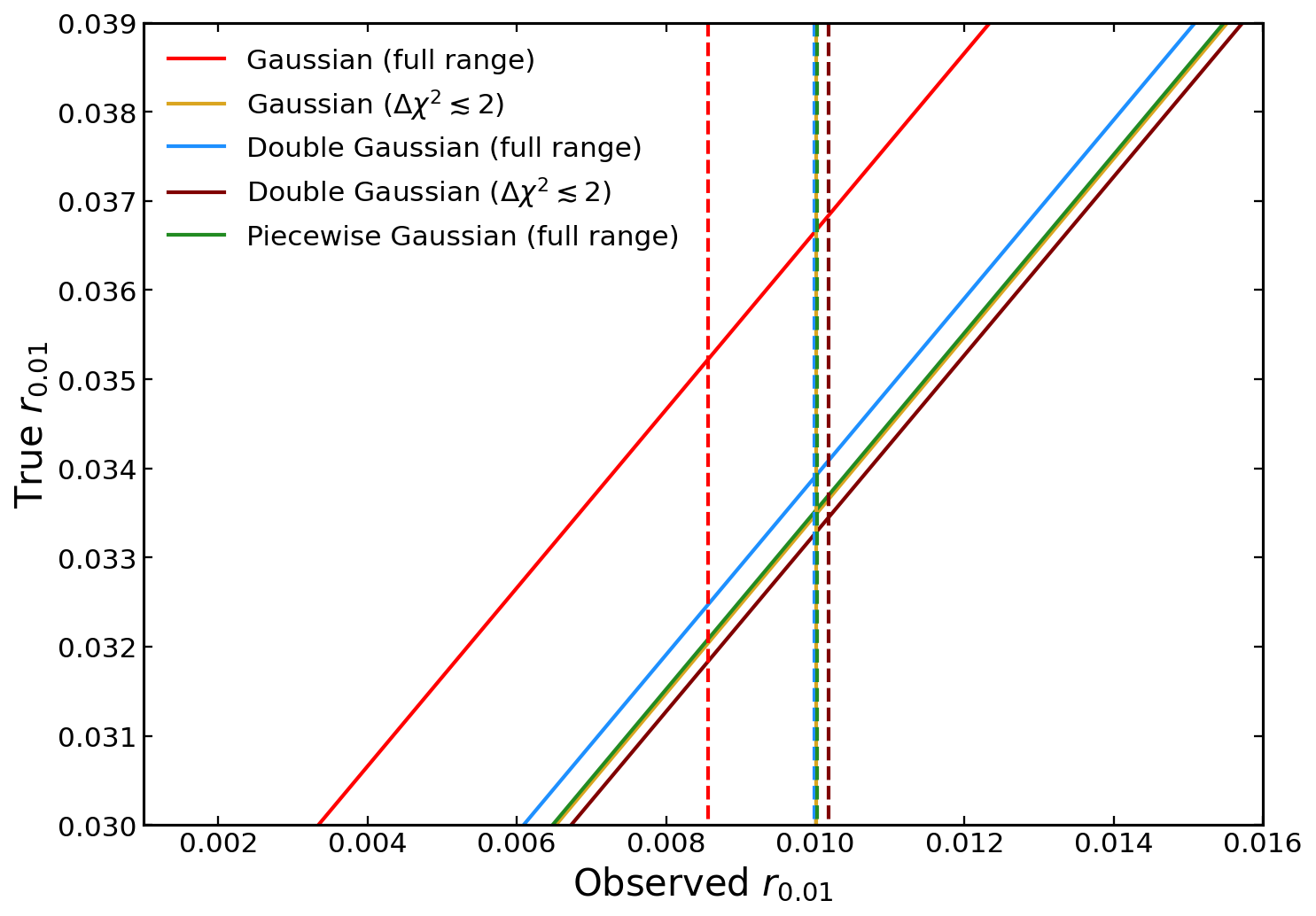}
    \caption{\textit{Left:} Fits of different curves on the PLK20(HiLLiPoP) + BK18 + LVK21 PL. \textit{Right:} Intersection between the FC belts (solid) and their relative best-fit values of $r_{0.01}$ (dashed), defining its upper bound with different fitting curves.}
    \label{fig:different_fits}
\end{figure*}

As an example, the left panel of \autoref{fig:different_fits} shows the results of PLK20(HiLLiPoP) + BK18 + LVK21, which is our most complex PL in terms of extra features. 
\begin{itemize}
\item Using the standard Gaussian with all the PL points, we clearly get a poor fit of the maximum likelihood because of the relatively high-probability tail. This can be solved by fitting just the points near the minimum ($\Delta\chi^2 \lesssim 2$), as we do in \autoref{sec:res_PL_r}.
\item If instead we try to use the double Gaussian on the full range of PL points, we can recover quite well both the maximum likelihood and the high-probability tail. On the other hand, using it in the region where $\Delta\chi^2 \lesssim 2$ allows us to fit the maximum and the bump at low $r_{0.01}$.
\item Similarly to the double Gaussian, using the piecewise Gaussian on the full range of points allows us to fit the maximum likelihood and the tail.
\end{itemize}

At this point, we perform the FC computation for all the cases mentioned above. The right panel of \autoref{fig:different_fits} shows the intersections between the various FC belts (solid) and their relative best-fit values for $r_{0.01}$ (dashed). We find that the upper bound of $r_{0.01}$ does not change significantly from case to case, allowing the choice made in \autoref{sec:res_PL_r}. Of course, the only relevant exception is the Gaussian case on the full range, which clearly overestimates the upper bound.

\bibliographystyle{apsrev}
\bibliography{bibliography}

\begin{thebibliography}{64}
\expandafter\ifx\csname natexlab\endcsname\relax\def\natexlab#1{#1}\fi
\expandafter\ifx\csname bibnamefont\endcsname\relax
  \def\bibnamefont#1{#1}\fi
\expandafter\ifx\csname bibfnamefont\endcsname\relax
  \def\bibfnamefont#1{#1}\fi
\expandafter\ifx\csname citenamefont\endcsname\relax
  \def\citenamefont#1{#1}\fi
\expandafter\ifx\csname url\endcsname\relax
  \def\url#1{\texttt{#1}}\fi
\expandafter\ifx\csname urlprefix\endcsname\relax\def\urlprefix{URL }\fi
\providecommand{\bibinfo}[2]{#2}
\providecommand{\eprint}[2][]{\url{#2}}

\bibitem[{\citenamefont{Seljak and Zaldarriaga}(1997)}]{Seljak_1997}
\bibinfo{author}{\bibfnamefont{U.}~\bibnamefont{Seljak}} \bibnamefont{and} \bibinfo{author}{\bibfnamefont{M.}~\bibnamefont{Zaldarriaga}}, \bibinfo{journal}{Physical Review Letters} \textbf{\bibinfo{volume}{78}}, \bibinfo{pages}{2054} (\bibinfo{year}{1997}), \eprint{astro-ph/9609169}.

\bibitem[{\citenamefont{Kamionkowski et~al.}(1997)\citenamefont{Kamionkowski, Kosowsky, and Stebbins}}]{Kamionkowski_1997}
\bibinfo{author}{\bibfnamefont{M.}~\bibnamefont{Kamionkowski}}, \bibinfo{author}{\bibfnamefont{A.}~\bibnamefont{Kosowsky}}, \bibnamefont{and} \bibinfo{author}{\bibfnamefont{A.}~\bibnamefont{Stebbins}}, \bibinfo{journal}{Physical Review D} \textbf{\bibinfo{volume}{55}}, \bibinfo{pages}{7368} (\bibinfo{year}{1997}), \eprint{astro-ph/9611125}.

\bibitem[{\citenamefont{Kamionkowski and Kovetz}(2016)}]{Kamionkowski_2016}
\bibinfo{author}{\bibfnamefont{M.}~\bibnamefont{Kamionkowski}} \bibnamefont{and} \bibinfo{author}{\bibfnamefont{E.~D.} \bibnamefont{Kovetz}}, \bibinfo{journal}{Annual Review of Astronomy and Astrophysics} \textbf{\bibinfo{volume}{54}}, \bibinfo{pages}{227} (\bibinfo{year}{2016}), \eprint{1510.06042}.

\bibitem[{\citenamefont{Guzzetti et~al.}(2016)\citenamefont{Guzzetti, Bartolo, Liguori, and Matarrese}}]{Guzzetti:2016mkm}
\bibinfo{author}{\bibfnamefont{M.~C.} \bibnamefont{Guzzetti}}, \bibinfo{author}{\bibfnamefont{N.}~\bibnamefont{Bartolo}}, \bibinfo{author}{\bibfnamefont{M.}~\bibnamefont{Liguori}}, \bibnamefont{and} \bibinfo{author}{\bibfnamefont{S.}~\bibnamefont{Matarrese}}, \bibinfo{journal}{Riv. Nuovo Cim.} \textbf{\bibinfo{volume}{39}}, \bibinfo{pages}{399} (\bibinfo{year}{2016}), \eprint{1605.01615}.

\bibitem[{\citenamefont{Seljak}(1997)}]{Seljak:1996ti}
\bibinfo{author}{\bibfnamefont{U.}~\bibnamefont{Seljak}}, \bibinfo{journal}{Astrophys. J.} \textbf{\bibinfo{volume}{482}}, \bibinfo{pages}{6} (\bibinfo{year}{1997}), \eprint{astro-ph/9608131}.

\bibitem[{\citenamefont{Hu and White}(1997)}]{Hu:1997hv}
\bibinfo{author}{\bibfnamefont{W.}~\bibnamefont{Hu}} \bibnamefont{and} \bibinfo{author}{\bibfnamefont{M.~J.} \bibnamefont{White}}, \bibinfo{journal}{New Astron.} \textbf{\bibinfo{volume}{2}}, \bibinfo{pages}{323} (\bibinfo{year}{1997}), \eprint{astro-ph/9706147}.

\bibitem[{\citenamefont{{BICEP2 Collaboration} et~al.}(2014)\citenamefont{{BICEP2 Collaboration}, {Ade}, {Aikin}, {Barkats}, {Benton}, {Bischoff}, {Bock}, {Brevik}, {Buder}, {Bullock} et~al.}}]{Ade_2014}
\bibinfo{author}{\bibnamefont{{BICEP2 Collaboration}}}, \bibinfo{author}{\bibfnamefont{P.~A.~R.} \bibnamefont{{Ade}}}, \bibinfo{author}{\bibfnamefont{R.~W.} \bibnamefont{{Aikin}}}, \bibinfo{author}{\bibfnamefont{D.}~\bibnamefont{{Barkats}}}, \bibinfo{author}{\bibfnamefont{S.~J.} \bibnamefont{{Benton}}}, \bibinfo{author}{\bibfnamefont{C.~A.} \bibnamefont{{Bischoff}}}, \bibinfo{author}{\bibfnamefont{J.~J.} \bibnamefont{{Bock}}}, \bibinfo{author}{\bibfnamefont{J.~A.} \bibnamefont{{Brevik}}}, \bibinfo{author}{\bibfnamefont{I.}~\bibnamefont{{Buder}}}, \bibinfo{author}{\bibfnamefont{E.}~\bibnamefont{{Bullock}}}, \bibnamefont{et~al.}, \bibinfo{journal}{\prl} \textbf{\bibinfo{volume}{112}}, \bibinfo{eid}{241101} (\bibinfo{year}{2014}), \eprint{1403.3985}.

\bibitem[{\citenamefont{{Simons Observatory Collaboration} et~al.}(2019)\citenamefont{{Simons Observatory Collaboration}, Ade, Aguirre, Ahmed, Aiola, Ali, Alonso, Alvarez, Arnold, Ashton et~al.}}]{Ade_2019}
\bibinfo{author}{\bibnamefont{{Simons Observatory Collaboration}}}, \bibinfo{author}{\bibfnamefont{P.}~\bibnamefont{Ade}}, \bibinfo{author}{\bibfnamefont{J.}~\bibnamefont{Aguirre}}, \bibinfo{author}{\bibfnamefont{Z.}~\bibnamefont{Ahmed}}, \bibinfo{author}{\bibfnamefont{S.}~\bibnamefont{Aiola}}, \bibinfo{author}{\bibfnamefont{A.}~\bibnamefont{Ali}}, \bibinfo{author}{\bibfnamefont{D.}~\bibnamefont{Alonso}}, \bibinfo{author}{\bibfnamefont{M.~A.} \bibnamefont{Alvarez}}, \bibinfo{author}{\bibfnamefont{K.}~\bibnamefont{Arnold}}, \bibinfo{author}{\bibfnamefont{P.}~\bibnamefont{Ashton}}, \bibnamefont{et~al.}, \bibinfo{journal}{Journal of Cosmology and Astroparticle Physics} \textbf{\bibinfo{volume}{2019}}, \bibinfo{pages}{056–056} (\bibinfo{year}{2019}), ISSN \bibinfo{issn}{1475-7516}, \eprint{1808.07445}.

\bibitem[{\citenamefont{Abazajian et~al.}(2016)\citenamefont{Abazajian, Adshead, Ahmed, Allen, Alonso, Arnold, Baccigalupi, Bartlett, Battaglia, Benson et~al.}}]{abazajian2016cmbs4}
\bibinfo{author}{\bibfnamefont{K.~N.} \bibnamefont{Abazajian}}, \bibinfo{author}{\bibfnamefont{P.}~\bibnamefont{Adshead}}, \bibinfo{author}{\bibfnamefont{Z.}~\bibnamefont{Ahmed}}, \bibinfo{author}{\bibfnamefont{S.~W.} \bibnamefont{Allen}}, \bibinfo{author}{\bibfnamefont{D.}~\bibnamefont{Alonso}}, \bibinfo{author}{\bibfnamefont{K.~S.} \bibnamefont{Arnold}}, \bibinfo{author}{\bibfnamefont{C.}~\bibnamefont{Baccigalupi}}, \bibinfo{author}{\bibfnamefont{J.~G.} \bibnamefont{Bartlett}}, \bibinfo{author}{\bibfnamefont{N.}~\bibnamefont{Battaglia}}, \bibinfo{author}{\bibfnamefont{B.~A.} \bibnamefont{Benson}}, \bibnamefont{et~al.} (\bibinfo{year}{2016}), \eprint{1610.02743}.

\bibitem[{\citenamefont{Hazumi et~al.}(2019)}]{Hazumi:2019lys}
\bibinfo{author}{\bibfnamefont{M.}~\bibnamefont{Hazumi}} \bibnamefont{et~al.}, \bibinfo{journal}{J. Low Temp. Phys.} \textbf{\bibinfo{volume}{194}}, \bibinfo{pages}{443} (\bibinfo{year}{2019}).

\bibitem[{\citenamefont{{LiteBIRD Collaboration} et~al.}(2022)\citenamefont{{LiteBIRD Collaboration}, Allys, Arnold, Aumont, Aurlien, Azzoni, Baccigalupi, Banday, Banerji, Barreiro et~al.}}]{PTEP_LiteBIRD}
\bibinfo{author}{\bibnamefont{{LiteBIRD Collaboration}}}, \bibinfo{author}{\bibfnamefont{E.}~\bibnamefont{Allys}}, \bibinfo{author}{\bibfnamefont{K.}~\bibnamefont{Arnold}}, \bibinfo{author}{\bibfnamefont{J.}~\bibnamefont{Aumont}}, \bibinfo{author}{\bibfnamefont{R.}~\bibnamefont{Aurlien}}, \bibinfo{author}{\bibfnamefont{S.}~\bibnamefont{Azzoni}}, \bibinfo{author}{\bibfnamefont{C.}~\bibnamefont{Baccigalupi}}, \bibinfo{author}{\bibfnamefont{A.~J.} \bibnamefont{Banday}}, \bibinfo{author}{\bibfnamefont{R.}~\bibnamefont{Banerji}}, \bibinfo{author}{\bibfnamefont{R.~B.} \bibnamefont{Barreiro}}, \bibnamefont{et~al.} (\bibinfo{year}{2022}), \eprint{2202.02773}.

\bibitem[{\citenamefont{{Planck Collaboration} et~al.}(2020{\natexlab{a}})\citenamefont{{Planck Collaboration}, Aghanim, Akrami, Ashdown, Aumont, Baccigalupi, Ballardini, Banday, Barreiro, Bartolo et~al.}}]{Planck_parameters}
\bibinfo{author}{\bibnamefont{{Planck Collaboration}}}, \bibinfo{author}{\bibfnamefont{N.}~\bibnamefont{Aghanim}}, \bibinfo{author}{\bibfnamefont{Y.}~\bibnamefont{Akrami}}, \bibinfo{author}{\bibfnamefont{M.}~\bibnamefont{Ashdown}}, \bibinfo{author}{\bibfnamefont{J.}~\bibnamefont{Aumont}}, \bibinfo{author}{\bibfnamefont{C.}~\bibnamefont{Baccigalupi}}, \bibinfo{author}{\bibfnamefont{M.}~\bibnamefont{Ballardini}}, \bibinfo{author}{\bibfnamefont{A.~J.} \bibnamefont{Banday}}, \bibinfo{author}{\bibfnamefont{R.~B.} \bibnamefont{Barreiro}}, \bibinfo{author}{\bibfnamefont{N.}~\bibnamefont{Bartolo}}, \bibnamefont{et~al.}, \bibinfo{journal}{Astron. Astrophys.} \textbf{\bibinfo{volume}{641}}, \bibinfo{pages}{A6} (\bibinfo{year}{2020}{\natexlab{a}}), \eprint{1807.06209}.

\bibitem[{\citenamefont{{Planck Collaboration} et~al.}(2020{\natexlab{b}})\citenamefont{{Planck Collaboration}, Aghanim, Akrami, Ashdown, Aumont, Baccigalupi, Ballardini, Banday, Barreiro, Bartolo et~al.}}]{Planck_like}
\bibinfo{author}{\bibnamefont{{Planck Collaboration}}}, \bibinfo{author}{\bibfnamefont{N.}~\bibnamefont{Aghanim}}, \bibinfo{author}{\bibfnamefont{Y.}~\bibnamefont{Akrami}}, \bibinfo{author}{\bibfnamefont{M.}~\bibnamefont{Ashdown}}, \bibinfo{author}{\bibfnamefont{J.}~\bibnamefont{Aumont}}, \bibinfo{author}{\bibfnamefont{C.}~\bibnamefont{Baccigalupi}}, \bibinfo{author}{\bibfnamefont{M.}~\bibnamefont{Ballardini}}, \bibinfo{author}{\bibfnamefont{A.~J.} \bibnamefont{Banday}}, \bibinfo{author}{\bibfnamefont{R.~B.} \bibnamefont{Barreiro}}, \bibinfo{author}{\bibfnamefont{N.}~\bibnamefont{Bartolo}}, \bibnamefont{et~al.}, \bibinfo{journal}{Astron. Astrophys.} \textbf{\bibinfo{volume}{641}}, \bibinfo{pages}{A5} (\bibinfo{year}{2020}{\natexlab{b}}), \eprint{1907.12875}.

\bibitem[{\citenamefont{{Rosenberg} et~al.}(2022)\citenamefont{{Rosenberg}, {Gratton}, and {Efstathiou}}}]{Rosenberg:2022}
\bibinfo{author}{\bibfnamefont{E.}~\bibnamefont{{Rosenberg}}}, \bibinfo{author}{\bibfnamefont{S.}~\bibnamefont{{Gratton}}}, \bibnamefont{and} \bibinfo{author}{\bibfnamefont{G.}~\bibnamefont{{Efstathiou}}}, \bibinfo{journal}{\mnras} \textbf{\bibinfo{volume}{517}}, \bibinfo{pages}{4620} (\bibinfo{year}{2022}), \eprint{2205.10869}.

\bibitem[{\citenamefont{{Tristram} et~al.}(2024)\citenamefont{{Tristram}, {Banday}, {Douspis}, {Garrido}, {G{\'o}rski}, {Henrot-Versill{\'e}}, {Hergt}, {Ili{\'c}}, {Keskitalo}, {Lagache} et~al.}}]{Tristram:2023}
\bibinfo{author}{\bibfnamefont{M.}~\bibnamefont{{Tristram}}}, \bibinfo{author}{\bibfnamefont{A.~J.} \bibnamefont{{Banday}}}, \bibinfo{author}{\bibfnamefont{M.}~\bibnamefont{{Douspis}}}, \bibinfo{author}{\bibfnamefont{X.}~\bibnamefont{{Garrido}}}, \bibinfo{author}{\bibfnamefont{K.~M.} \bibnamefont{{G{\'o}rski}}}, \bibinfo{author}{\bibfnamefont{S.}~\bibnamefont{{Henrot-Versill{\'e}}}}, \bibinfo{author}{\bibfnamefont{L.~T.} \bibnamefont{{Hergt}}}, \bibinfo{author}{\bibfnamefont{S.}~\bibnamefont{{Ili{\'c}}}}, \bibinfo{author}{\bibfnamefont{R.}~\bibnamefont{{Keskitalo}}}, \bibinfo{author}{\bibfnamefont{G.}~\bibnamefont{{Lagache}}}, \bibnamefont{et~al.}, \bibinfo{journal}{\aap} \textbf{\bibinfo{volume}{682}}, \bibinfo{eid}{A37} (\bibinfo{year}{2024}), \eprint{2309.10034}.

\bibitem[{\citenamefont{Tristram et~al.}(2022)\citenamefont{Tristram, Banday, G\'orski, Keskitalo, Lawrence, Andersen, Barreiro, Borrill, Colombo, Eriksen et~al.}}]{Tristram:2022}
\bibinfo{author}{\bibfnamefont{M.}~\bibnamefont{Tristram}}, \bibinfo{author}{\bibfnamefont{A.~J.} \bibnamefont{Banday}}, \bibinfo{author}{\bibfnamefont{K.~M.} \bibnamefont{G\'orski}}, \bibinfo{author}{\bibfnamefont{R.}~\bibnamefont{Keskitalo}}, \bibinfo{author}{\bibfnamefont{C.~R.} \bibnamefont{Lawrence}}, \bibinfo{author}{\bibfnamefont{K.~J.} \bibnamefont{Andersen}}, \bibinfo{author}{\bibfnamefont{R.~B.} \bibnamefont{Barreiro}}, \bibinfo{author}{\bibfnamefont{J.}~\bibnamefont{Borrill}}, \bibinfo{author}{\bibfnamefont{L.~P.~L.} \bibnamefont{Colombo}}, \bibinfo{author}{\bibfnamefont{H.~K.} \bibnamefont{Eriksen}}, \bibnamefont{et~al.}, \bibinfo{journal}{Phys. Rev. D} \textbf{\bibinfo{volume}{105}}, \bibinfo{pages}{083524} (\bibinfo{year}{2022}).

\bibitem[{\citenamefont{{Galloni} et~al.}(2023)\citenamefont{{Galloni}, {Bartolo}, {Matarrese}, {Migliaccio}, {Ricciardone}, and {Vittorio}}}]{Galloni:2023}
\bibinfo{author}{\bibfnamefont{G.}~\bibnamefont{{Galloni}}}, \bibinfo{author}{\bibfnamefont{N.}~\bibnamefont{{Bartolo}}}, \bibinfo{author}{\bibfnamefont{S.}~\bibnamefont{{Matarrese}}}, \bibinfo{author}{\bibfnamefont{M.}~\bibnamefont{{Migliaccio}}}, \bibinfo{author}{\bibfnamefont{A.}~\bibnamefont{{Ricciardone}}}, \bibnamefont{and} \bibinfo{author}{\bibfnamefont{N.}~\bibnamefont{{Vittorio}}}, \bibinfo{journal}{\jcap} \textbf{\bibinfo{volume}{2023}}, \bibinfo{eid}{062} (\bibinfo{year}{2023}), \eprint{2208.00188}.

\bibitem[{\citenamefont{{Planck Collaboration} et~al.}(2020{\natexlab{c}})\citenamefont{{Planck Collaboration}, Akrami, Arroja, Ashdown, Aumont, Baccigalupi, Ballardini, Banday, Barreiro, Bartolo et~al.}}]{Planck_2018}
\bibinfo{author}{\bibnamefont{{Planck Collaboration}}}, \bibinfo{author}{\bibfnamefont{Y.}~\bibnamefont{Akrami}}, \bibinfo{author}{\bibfnamefont{F.}~\bibnamefont{Arroja}}, \bibinfo{author}{\bibfnamefont{M.}~\bibnamefont{Ashdown}}, \bibinfo{author}{\bibfnamefont{J.}~\bibnamefont{Aumont}}, \bibinfo{author}{\bibfnamefont{C.}~\bibnamefont{Baccigalupi}}, \bibinfo{author}{\bibfnamefont{M.}~\bibnamefont{Ballardini}}, \bibinfo{author}{\bibfnamefont{A.~J.} \bibnamefont{Banday}}, \bibinfo{author}{\bibfnamefont{R.~B.} \bibnamefont{Barreiro}}, \bibinfo{author}{\bibfnamefont{N.}~\bibnamefont{Bartolo}}, \bibnamefont{et~al.}, \bibinfo{journal}{Astronomy \& Astrophysics} \textbf{\bibinfo{volume}{641}}, \bibinfo{pages}{A10} (\bibinfo{year}{2020}{\natexlab{c}}), ISSN \bibinfo{issn}{1432-0746}, \eprint{1807.06211}.

\bibitem[{\citenamefont{{LIGO Scientific Collaboration} et~al.}(2015)\citenamefont{{LIGO Scientific Collaboration}, Aasi, Abbott, Abbott, Abbott, Abernathy, Ackley, Adams, Adams, Addesso et~al.}}]{aLIGO}
\bibinfo{author}{\bibnamefont{{LIGO Scientific Collaboration}}}, \bibinfo{author}{\bibfnamefont{J.}~\bibnamefont{Aasi}}, \bibinfo{author}{\bibfnamefont{B.~P.} \bibnamefont{Abbott}}, \bibinfo{author}{\bibfnamefont{R.}~\bibnamefont{Abbott}}, \bibinfo{author}{\bibfnamefont{T.}~\bibnamefont{Abbott}}, \bibinfo{author}{\bibfnamefont{M.~R.} \bibnamefont{Abernathy}}, \bibinfo{author}{\bibfnamefont{K.}~\bibnamefont{Ackley}}, \bibinfo{author}{\bibfnamefont{C.}~\bibnamefont{Adams}}, \bibinfo{author}{\bibfnamefont{T.}~\bibnamefont{Adams}}, \bibinfo{author}{\bibfnamefont{P.}~\bibnamefont{Addesso}}, \bibnamefont{et~al.}, \bibinfo{journal}{Classical and Quantum Gravity} \textbf{\bibinfo{volume}{32}}, \bibinfo{pages}{074001} (\bibinfo{year}{2015}), ISSN \bibinfo{issn}{1361-6382}, \eprint{1411.4547}.

\bibitem[{\citenamefont{Acernese et~al.}(2014)\citenamefont{Acernese, Agathos, Agatsuma, Aisa, Allemandou, Allocca, Amarni, Astone, Balestri, Ballardin et~al.}}]{aVirgo}
\bibinfo{author}{\bibfnamefont{F.}~\bibnamefont{Acernese}}, \bibinfo{author}{\bibfnamefont{M.}~\bibnamefont{Agathos}}, \bibinfo{author}{\bibfnamefont{K.}~\bibnamefont{Agatsuma}}, \bibinfo{author}{\bibfnamefont{D.}~\bibnamefont{Aisa}}, \bibinfo{author}{\bibfnamefont{N.}~\bibnamefont{Allemandou}}, \bibinfo{author}{\bibfnamefont{A.}~\bibnamefont{Allocca}}, \bibinfo{author}{\bibfnamefont{J.}~\bibnamefont{Amarni}}, \bibinfo{author}{\bibfnamefont{P.}~\bibnamefont{Astone}}, \bibinfo{author}{\bibfnamefont{G.}~\bibnamefont{Balestri}}, \bibinfo{author}{\bibfnamefont{G.}~\bibnamefont{Ballardin}}, \bibnamefont{et~al.}, \bibinfo{journal}{Classical and Quantum Gravity} \textbf{\bibinfo{volume}{32}}, \bibinfo{pages}{024001} (\bibinfo{year}{2014}), ISSN \bibinfo{issn}{1361-6382}, \eprint{1408.3978}.

\bibitem[{\citenamefont{{LIGO Scientific Collaboration} et~al.}(2020)\citenamefont{{LIGO Scientific Collaboration}, {VIRGO Collaboration}, {Kagra Collaboration}, {Abbott}, {Abbott}, {Abbott}, {Abraham}, {Acernese}, {Ackley}, {Adams} et~al.}}]{ligo-virgo}
\bibinfo{author}{\bibnamefont{{LIGO Scientific Collaboration}}}, \bibinfo{author}{\bibnamefont{{VIRGO Collaboration}}}, \bibinfo{author}{\bibnamefont{{Kagra Collaboration}}}, \bibinfo{author}{\bibfnamefont{B.~P.} \bibnamefont{{Abbott}}}, \bibinfo{author}{\bibfnamefont{R.}~\bibnamefont{{Abbott}}}, \bibinfo{author}{\bibfnamefont{T.~D.} \bibnamefont{{Abbott}}}, \bibinfo{author}{\bibfnamefont{S.}~\bibnamefont{{Abraham}}}, \bibinfo{author}{\bibfnamefont{F.}~\bibnamefont{{Acernese}}}, \bibinfo{author}{\bibfnamefont{K.}~\bibnamefont{{Ackley}}}, \bibinfo{author}{\bibfnamefont{C.}~\bibnamefont{{Adams}}}, \bibnamefont{et~al.}, \bibinfo{journal}{Living Reviews in Relativity} \textbf{\bibinfo{volume}{23}}, \bibinfo{eid}{3} (\bibinfo{year}{2020}).

\bibitem[{\citenamefont{{LIGO Scientific} et~al.}(2019)\citenamefont{{LIGO Scientific}, {Virgo Collaboration}, Abbott, Abbott, Abbott, Abraham, Acernese, Ackley, Adams, Adya et~al.}}]{Abbott_2019_ligo}
\bibinfo{author}{\bibnamefont{{LIGO Scientific}}}, \bibinfo{author}{\bibnamefont{{Virgo Collaboration}}}, \bibinfo{author}{\bibfnamefont{B.~P.} \bibnamefont{Abbott}}, \bibinfo{author}{\bibfnamefont{R.}~\bibnamefont{Abbott}}, \bibinfo{author}{\bibfnamefont{T.~D.} \bibnamefont{Abbott}}, \bibinfo{author}{\bibfnamefont{S.}~\bibnamefont{Abraham}}, \bibinfo{author}{\bibfnamefont{F.}~\bibnamefont{Acernese}}, \bibinfo{author}{\bibfnamefont{K.}~\bibnamefont{Ackley}}, \bibinfo{author}{\bibfnamefont{C.}~\bibnamefont{Adams}}, \bibinfo{author}{\bibfnamefont{V.~B.} \bibnamefont{Adya}}, \bibnamefont{et~al.}, \bibinfo{journal}{Phys. Rev. D} \textbf{\bibinfo{volume}{100}}, \bibinfo{pages}{061101} (\bibinfo{year}{2019}).

\bibitem[{\citenamefont{{LIGO Scientific Collaboration} et~al.}(2021)\citenamefont{{LIGO Scientific Collaboration}, {Virgo Collaboration}, {KAGRA Collaboration}, Abbott, Abbott, Abraham, Acernese, Ackley, Adams, Adams et~al.}}]{ligo_meas}
\bibinfo{author}{\bibnamefont{{LIGO Scientific Collaboration}}}, \bibinfo{author}{\bibnamefont{{Virgo Collaboration}}}, \bibinfo{author}{\bibnamefont{{KAGRA Collaboration}}}, \bibinfo{author}{\bibfnamefont{R.}~\bibnamefont{Abbott}}, \bibinfo{author}{\bibfnamefont{T.~D.} \bibnamefont{Abbott}}, \bibinfo{author}{\bibfnamefont{S.}~\bibnamefont{Abraham}}, \bibinfo{author}{\bibfnamefont{F.}~\bibnamefont{Acernese}}, \bibinfo{author}{\bibfnamefont{K.}~\bibnamefont{Ackley}}, \bibinfo{author}{\bibfnamefont{A.}~\bibnamefont{Adams}}, \bibinfo{author}{\bibfnamefont{C.}~\bibnamefont{Adams}}, \bibnamefont{et~al.}, \bibinfo{journal}{Phys. Rev. D} \textbf{\bibinfo{volume}{104}}, \bibinfo{pages}{022004} (\bibinfo{year}{2021}).

\bibitem[{\citenamefont{Metropolis and Ulam}(1949)}]{MCMethod}
\bibinfo{author}{\bibfnamefont{N.}~\bibnamefont{Metropolis}} \bibnamefont{and} \bibinfo{author}{\bibfnamefont{S.}~\bibnamefont{Ulam}}, \bibinfo{journal}{Journal of the American Statistical Association} \textbf{\bibinfo{volume}{44}}, \bibinfo{pages}{335} (\bibinfo{year}{1949}), ISSN \bibinfo{issn}{01621459}.

\bibitem[{\citenamefont{Eckhardt}(1987)}]{Ulam}
\bibinfo{author}{\bibfnamefont{R.}~\bibnamefont{Eckhardt}}, \bibinfo{journal}{Los Alamos Science} \textbf{\bibinfo{volume}{15}}, \bibinfo{pages}{131} (\bibinfo{year}{1987}).

\bibitem[{\citenamefont{Kroese et~al.}(2014)\citenamefont{Kroese, Brereton, Taimre, and Botev}}]{MonteCarlo}
\bibinfo{author}{\bibfnamefont{D.~P.} \bibnamefont{Kroese}}, \bibinfo{author}{\bibfnamefont{T.}~\bibnamefont{Brereton}}, \bibinfo{author}{\bibfnamefont{T.}~\bibnamefont{Taimre}}, \bibnamefont{and} \bibinfo{author}{\bibfnamefont{Z.~I.} \bibnamefont{Botev}}, \bibinfo{journal}{WIREs Computational Statistics} \textbf{\bibinfo{volume}{6}}, \bibinfo{pages}{386} (\bibinfo{year}{2014}).

\bibitem[{\citenamefont{Lewis and Bridle}(2002)}]{Lewis:2002ah}
\bibinfo{author}{\bibfnamefont{A.}~\bibnamefont{Lewis}} \bibnamefont{and} \bibinfo{author}{\bibfnamefont{S.}~\bibnamefont{Bridle}}, \bibinfo{journal}{Phys. Rev.} \textbf{\bibinfo{volume}{D66}}, \bibinfo{pages}{103511} (\bibinfo{year}{2002}), \eprint{astro-ph/0205436}.

\bibitem[{\citenamefont{Lewis}(2013)}]{Lewis:2013hha}
\bibinfo{author}{\bibfnamefont{A.}~\bibnamefont{Lewis}}, \bibinfo{journal}{Phys. Rev.} \textbf{\bibinfo{volume}{D87}}, \bibinfo{pages}{103529} (\bibinfo{year}{2013}), \eprint{1304.4473}.

\bibitem[{\citenamefont{Efstathiou et~al.}(2023)\citenamefont{Efstathiou, Rosenberg, and Poulin}}]{Efstathiou:2023fbn}
\bibinfo{author}{\bibfnamefont{G.}~\bibnamefont{Efstathiou}}, \bibinfo{author}{\bibfnamefont{E.}~\bibnamefont{Rosenberg}}, \bibnamefont{and} \bibinfo{author}{\bibfnamefont{V.}~\bibnamefont{Poulin}} (\bibinfo{year}{2023}), \eprint{2311.00524}.

\bibitem[{\citenamefont{Henrot-Versillé et~al.}(2015)\citenamefont{Henrot-Versillé, Robinet, Leroy, Plaszczynski, Arnaud, Bizouard, Cavalier, Christensen, Couchot, Franco et~al.}}]{Henrot_Versill__2015}
\bibinfo{author}{\bibfnamefont{S.}~\bibnamefont{Henrot-Versillé}}, \bibinfo{author}{\bibfnamefont{F.}~\bibnamefont{Robinet}}, \bibinfo{author}{\bibfnamefont{N.}~\bibnamefont{Leroy}}, \bibinfo{author}{\bibfnamefont{S.}~\bibnamefont{Plaszczynski}}, \bibinfo{author}{\bibfnamefont{N.}~\bibnamefont{Arnaud}}, \bibinfo{author}{\bibfnamefont{M.-A.} \bibnamefont{Bizouard}}, \bibinfo{author}{\bibfnamefont{F.}~\bibnamefont{Cavalier}}, \bibinfo{author}{\bibfnamefont{N.}~\bibnamefont{Christensen}}, \bibinfo{author}{\bibfnamefont{F.}~\bibnamefont{Couchot}}, \bibinfo{author}{\bibfnamefont{S.}~\bibnamefont{Franco}}, \bibnamefont{et~al.}, \bibinfo{journal}{Classical and Quantum Gravity} \textbf{\bibinfo{volume}{32}}, \bibinfo{pages}{045003} (\bibinfo{year}{2015}), ISSN \bibinfo{issn}{1361-6382}.

\bibitem[{\citenamefont{Henrot-Versill\'e et~al.}(2016)\citenamefont{Henrot-Versill\'e, Perdereau, Plaszczynski, d'Orfeuil, Spinelli, and Tristram}}]{Henrot-Versille:2016htt}
\bibinfo{author}{\bibfnamefont{S.}~\bibnamefont{Henrot-Versill\'e}}, \bibinfo{author}{\bibfnamefont{O.}~\bibnamefont{Perdereau}}, \bibinfo{author}{\bibfnamefont{S.}~\bibnamefont{Plaszczynski}}, \bibinfo{author}{\bibfnamefont{B.~R.} \bibnamefont{d'Orfeuil}}, \bibinfo{author}{\bibfnamefont{M.}~\bibnamefont{Spinelli}}, \bibnamefont{and} \bibinfo{author}{\bibfnamefont{M.}~\bibnamefont{Tristram}} (\bibinfo{year}{2016}), \eprint{1607.02964}.

\bibitem[{\citenamefont{Herold et~al.}(2022)\citenamefont{Herold, Ferreira, and Komatsu}}]{Herold_2022}
\bibinfo{author}{\bibfnamefont{L.}~\bibnamefont{Herold}}, \bibinfo{author}{\bibfnamefont{E.~G.~M.} \bibnamefont{Ferreira}}, \bibnamefont{and} \bibinfo{author}{\bibfnamefont{E.}~\bibnamefont{Komatsu}}, \bibinfo{journal}{The Astrophysical Journal Letters} \textbf{\bibinfo{volume}{929}}, \bibinfo{pages}{L16} (\bibinfo{year}{2022}), ISSN \bibinfo{issn}{2041-8213}.

\bibitem[{\citenamefont{Murgia et~al.}(2021)\citenamefont{Murgia, Abell\'an, and Poulin}}]{Murgia_2021}
\bibinfo{author}{\bibfnamefont{R.}~\bibnamefont{Murgia}}, \bibinfo{author}{\bibfnamefont{G.~F.} \bibnamefont{Abell\'an}}, \bibnamefont{and} \bibinfo{author}{\bibfnamefont{V.}~\bibnamefont{Poulin}}, \bibinfo{journal}{Phys. Rev. D} \textbf{\bibinfo{volume}{103}}, \bibinfo{pages}{063502} (\bibinfo{year}{2021}).

\bibitem[{\citenamefont{McDonough et~al.}(2023)\citenamefont{McDonough, Hill, Ivanov, La~Posta, and Toomey}}]{McDonough:2023qcu}
\bibinfo{author}{\bibfnamefont{E.}~\bibnamefont{McDonough}}, \bibinfo{author}{\bibfnamefont{J.~C.} \bibnamefont{Hill}}, \bibinfo{author}{\bibfnamefont{M.~M.} \bibnamefont{Ivanov}}, \bibinfo{author}{\bibfnamefont{A.}~\bibnamefont{La~Posta}}, \bibnamefont{and} \bibinfo{author}{\bibfnamefont{M.~W.} \bibnamefont{Toomey}} (\bibinfo{year}{2023}), \eprint{2310.19899}.

\bibitem[{\citenamefont{{Planck Collaboration} et~al.}(2020{\natexlab{d}})\citenamefont{{Planck Collaboration}, {Akrami}, {Andersen}, {Ashdown}, {Baccigalupi}, {Ballardini}, {Banday}, {Barreiro}, {Bartolo}, {Basak} et~al.}}]{planck2020-LVII}
\bibinfo{author}{\bibnamefont{{Planck Collaboration}}}, \bibinfo{author}{\bibfnamefont{Y.}~\bibnamefont{{Akrami}}}, \bibinfo{author}{\bibfnamefont{K.~J.} \bibnamefont{{Andersen}}}, \bibinfo{author}{\bibfnamefont{M.}~\bibnamefont{{Ashdown}}}, \bibinfo{author}{\bibfnamefont{C.}~\bibnamefont{{Baccigalupi}}}, \bibinfo{author}{\bibfnamefont{M.}~\bibnamefont{{Ballardini}}}, \bibinfo{author}{\bibfnamefont{A.~J.} \bibnamefont{{Banday}}}, \bibinfo{author}{\bibfnamefont{R.~B.} \bibnamefont{{Barreiro}}}, \bibinfo{author}{\bibfnamefont{N.}~\bibnamefont{{Bartolo}}}, \bibinfo{author}{\bibfnamefont{S.}~\bibnamefont{{Basak}}}, \bibnamefont{et~al.}, \bibinfo{journal}{\aap} \textbf{\bibinfo{volume}{643}}, \bibinfo{eid}{A42} (\bibinfo{year}{2020}{\natexlab{d}}), \eprint{2007.04997}.

\bibitem[{\citenamefont{{Eriksen} et~al.}(2008)\citenamefont{{Eriksen}, {Jewell}, {Dickinson}, {Banday}, {G{\'o}rski}, and {Lawrence}}}]{Eriksen_2008}
\bibinfo{author}{\bibfnamefont{H.~K.} \bibnamefont{{Eriksen}}}, \bibinfo{author}{\bibfnamefont{J.~B.} \bibnamefont{{Jewell}}}, \bibinfo{author}{\bibfnamefont{C.}~\bibnamefont{{Dickinson}}}, \bibinfo{author}{\bibfnamefont{A.~J.} \bibnamefont{{Banday}}}, \bibinfo{author}{\bibfnamefont{K.~M.} \bibnamefont{{G{\'o}rski}}}, \bibnamefont{and} \bibinfo{author}{\bibfnamefont{C.~R.} \bibnamefont{{Lawrence}}}, \bibinfo{journal}{\apj} \textbf{\bibinfo{volume}{676}}, \bibinfo{pages}{10} (\bibinfo{year}{2008}), \eprint{0709.1058}.

\bibitem[{\citenamefont{{BICEP/Keck Collaboration} et~al.}(2021)\citenamefont{{BICEP/Keck Collaboration}, Ade, Ahmed, Amiri, Barkats, Thakur, Bischoff, Beck, Bock, Boenish et~al.}}]{BICEP_2021}
\bibinfo{author}{\bibnamefont{{BICEP/Keck Collaboration}}}, \bibinfo{author}{\bibfnamefont{P.~A.~R.} \bibnamefont{Ade}}, \bibinfo{author}{\bibfnamefont{Z.}~\bibnamefont{Ahmed}}, \bibinfo{author}{\bibfnamefont{M.}~\bibnamefont{Amiri}}, \bibinfo{author}{\bibfnamefont{D.}~\bibnamefont{Barkats}}, \bibinfo{author}{\bibfnamefont{R.~B.} \bibnamefont{Thakur}}, \bibinfo{author}{\bibfnamefont{C.~A.} \bibnamefont{Bischoff}}, \bibinfo{author}{\bibfnamefont{D.}~\bibnamefont{Beck}}, \bibinfo{author}{\bibfnamefont{J.~J.} \bibnamefont{Bock}}, \bibinfo{author}{\bibfnamefont{H.}~\bibnamefont{Boenish}}, \bibnamefont{et~al.}, \bibinfo{journal}{Phys. Rev. Lett.} \textbf{\bibinfo{volume}{127}}, \bibinfo{pages}{151301} (\bibinfo{year}{2021}).

\bibitem[{\citenamefont{{Hamimeche} and {Lewis}}(2008)}]{Hamimeche_2008}
\bibinfo{author}{\bibfnamefont{S.}~\bibnamefont{{Hamimeche}}} \bibnamefont{and} \bibinfo{author}{\bibfnamefont{A.}~\bibnamefont{{Lewis}}}, \bibinfo{journal}{\prd} \textbf{\bibinfo{volume}{77}}, \bibinfo{eid}{103013} (\bibinfo{year}{2008}), \eprint{0801.0554}.

\bibitem[{\citenamefont{{Cabass} et~al.}(2016)\citenamefont{{Cabass}, {Pagano}, {Salvati}, {Gerbino}, {Giusarma}, and {Melchiorri}}}]{Cabass_2016}
\bibinfo{author}{\bibfnamefont{G.}~\bibnamefont{{Cabass}}}, \bibinfo{author}{\bibfnamefont{L.}~\bibnamefont{{Pagano}}}, \bibinfo{author}{\bibfnamefont{L.}~\bibnamefont{{Salvati}}}, \bibinfo{author}{\bibfnamefont{M.}~\bibnamefont{{Gerbino}}}, \bibinfo{author}{\bibfnamefont{E.}~\bibnamefont{{Giusarma}}}, \bibnamefont{and} \bibinfo{author}{\bibfnamefont{A.}~\bibnamefont{{Melchiorri}}}, \bibinfo{journal}{\prd} \textbf{\bibinfo{volume}{93}}, \bibinfo{eid}{063508} (\bibinfo{year}{2016}), \eprint{1511.05146}.

\bibitem[{\citenamefont{{Carron} et~al.}(2022)\citenamefont{{Carron}, {Mirmelstein}, and {Lewis}}}]{carron_2022}
\bibinfo{author}{\bibfnamefont{J.}~\bibnamefont{{Carron}}}, \bibinfo{author}{\bibfnamefont{M.}~\bibnamefont{{Mirmelstein}}}, \bibnamefont{and} \bibinfo{author}{\bibfnamefont{A.}~\bibnamefont{{Lewis}}}, \bibinfo{journal}{\jcap} \textbf{\bibinfo{volume}{2022}}, \bibinfo{eid}{039} (\bibinfo{year}{2022}), \eprint{2206.07773}.

\bibitem[{\citenamefont{Lewis et~al.}(2000)\citenamefont{Lewis, Challinor, and Lasenby}}]{Lewis_2000}
\bibinfo{author}{\bibfnamefont{A.}~\bibnamefont{Lewis}}, \bibinfo{author}{\bibfnamefont{A.}~\bibnamefont{Challinor}}, \bibnamefont{and} \bibinfo{author}{\bibfnamefont{A.}~\bibnamefont{Lasenby}}, \bibinfo{journal}{The Astrophysical Journal} \textbf{\bibinfo{volume}{538}}, \bibinfo{pages}{473} (\bibinfo{year}{2000}), \eprint{astro-ph/9911177}.

\bibitem[{\citenamefont{Howlett et~al.}(2012)\citenamefont{Howlett, Lewis, Hall, and Challinor}}]{Howlett_2012}
\bibinfo{author}{\bibfnamefont{C.}~\bibnamefont{Howlett}}, \bibinfo{author}{\bibfnamefont{A.}~\bibnamefont{Lewis}}, \bibinfo{author}{\bibfnamefont{A.}~\bibnamefont{Hall}}, \bibnamefont{and} \bibinfo{author}{\bibfnamefont{A.}~\bibnamefont{Challinor}}, \bibinfo{journal}{Journal of Cosmology and Astroparticle Physics} \textbf{\bibinfo{volume}{2012}}, \bibinfo{pages}{027} (\bibinfo{year}{2012}), \eprint{1201.3654}.

\bibitem[{\citenamefont{Couchot et~al.}(2017)\citenamefont{Couchot, Henrot-Versillé, Perdereau, Plaszczynski, Rouillé~d’Orfeuil, Spinelli, and Tristram}}]{Couchot_2017}
\bibinfo{author}{\bibfnamefont{F.}~\bibnamefont{Couchot}}, \bibinfo{author}{\bibfnamefont{S.}~\bibnamefont{Henrot-Versillé}}, \bibinfo{author}{\bibfnamefont{O.}~\bibnamefont{Perdereau}}, \bibinfo{author}{\bibfnamefont{S.}~\bibnamefont{Plaszczynski}}, \bibinfo{author}{\bibfnamefont{B.}~\bibnamefont{Rouillé~d’Orfeuil}}, \bibinfo{author}{\bibfnamefont{M.}~\bibnamefont{Spinelli}}, \bibnamefont{and} \bibinfo{author}{\bibfnamefont{M.}~\bibnamefont{Tristram}}, \bibinfo{journal}{Astronomy \& Astrophysics} \textbf{\bibinfo{volume}{606}}, \bibinfo{pages}{A104} (\bibinfo{year}{2017}), ISSN \bibinfo{issn}{1432-0746}.

\bibitem[{\citenamefont{Campeti and Komatsu}(2022)}]{Campeti:2022vom}
\bibinfo{author}{\bibfnamefont{P.}~\bibnamefont{Campeti}} \bibnamefont{and} \bibinfo{author}{\bibfnamefont{E.}~\bibnamefont{Komatsu}}, \bibinfo{journal}{Astrophys. J.} \textbf{\bibinfo{volume}{941}}, \bibinfo{pages}{110} (\bibinfo{year}{2022}), \eprint{2205.05617}.

\bibitem[{\citenamefont{Acquaviva et~al.}(2003)\citenamefont{Acquaviva, Bartolo, Matarrese, and Riotto}}]{acquaviva2003SecondOrderCosmologicalPerturbationsInflation}
\bibinfo{author}{\bibfnamefont{V.}~\bibnamefont{Acquaviva}}, \bibinfo{author}{\bibfnamefont{N.}~\bibnamefont{Bartolo}}, \bibinfo{author}{\bibfnamefont{S.}~\bibnamefont{Matarrese}}, \bibnamefont{and} \bibinfo{author}{\bibfnamefont{A.}~\bibnamefont{Riotto}}, \bibinfo{journal}{Nuclear Physics B} \textbf{\bibinfo{volume}{667}}, \bibinfo{pages}{119} (\bibinfo{year}{2003}), ISSN \bibinfo{issn}{05503213}, \eprint{astro-ph/0209156}.

\bibitem[{\citenamefont{Matarrese et~al.}(1998)\citenamefont{Matarrese, Mollerach, and Bruni}}]{matarrese1998RelativisticSecondorderPerturbationsEinsteinde}
\bibinfo{author}{\bibfnamefont{S.}~\bibnamefont{Matarrese}}, \bibinfo{author}{\bibfnamefont{S.}~\bibnamefont{Mollerach}}, \bibnamefont{and} \bibinfo{author}{\bibfnamefont{M.}~\bibnamefont{Bruni}}, \bibinfo{journal}{Physical Review D} \textbf{\bibinfo{volume}{58}}, \bibinfo{pages}{043504} (\bibinfo{year}{1998}), ISSN \bibinfo{issn}{0556-2821, 1089-4918}, \eprint{astro-ph/9707278}.

\bibitem[{\citenamefont{{Baumann} et~al.}(2007)\citenamefont{{Baumann}, {Steinhardt}, {Takahashi}, and {Ichiki}}}]{baumann2007GravitationalWaveSpectrumInduced}
\bibinfo{author}{\bibfnamefont{D.}~\bibnamefont{{Baumann}}}, \bibinfo{author}{\bibfnamefont{P.}~\bibnamefont{{Steinhardt}}}, \bibinfo{author}{\bibfnamefont{K.}~\bibnamefont{{Takahashi}}}, \bibnamefont{and} \bibinfo{author}{\bibfnamefont{K.}~\bibnamefont{{Ichiki}}}, \bibinfo{journal}{\prd} \textbf{\bibinfo{volume}{76}}, \bibinfo{eid}{084019} (\bibinfo{year}{2007}), \eprint{hep-th/0703290}.

\bibitem[{\citenamefont{Ananda et~al.}(2007)\citenamefont{Ananda, Clarkson, and Wands}}]{ananda2007CosmologicalGravitationalWaveBackground}
\bibinfo{author}{\bibfnamefont{K.~N.} \bibnamefont{Ananda}}, \bibinfo{author}{\bibfnamefont{C.}~\bibnamefont{Clarkson}}, \bibnamefont{and} \bibinfo{author}{\bibfnamefont{D.}~\bibnamefont{Wands}}, \bibinfo{journal}{Physical Review D} \textbf{\bibinfo{volume}{75}}, \bibinfo{pages}{123518} (\bibinfo{year}{2007}), ISSN \bibinfo{issn}{1550-7998, 1550-2368}, \eprint{gr-qc/0612013}.

\bibitem[{\citenamefont{Fidler et~al.}(2014)\citenamefont{Fidler, Pettinari, Beneke, Crittenden, Koyama, and Wands}}]{fidler2014IntrinsicBmodePolarisationCosmic}
\bibinfo{author}{\bibfnamefont{C.}~\bibnamefont{Fidler}}, \bibinfo{author}{\bibfnamefont{G.~W.} \bibnamefont{Pettinari}}, \bibinfo{author}{\bibfnamefont{M.}~\bibnamefont{Beneke}}, \bibinfo{author}{\bibfnamefont{R.}~\bibnamefont{Crittenden}}, \bibinfo{author}{\bibfnamefont{K.}~\bibnamefont{Koyama}}, \bibnamefont{and} \bibinfo{author}{\bibfnamefont{D.}~\bibnamefont{Wands}}, \bibinfo{journal}{Journal of Cosmology and Astroparticle Physics} \textbf{\bibinfo{volume}{07}}, \bibinfo{pages}{011} (\bibinfo{year}{2014}), ISSN \bibinfo{issn}{1475-7516}, \eprint{1401.3296}.

\bibitem[{\citenamefont{Hergt et~al.}(2021)\citenamefont{Hergt, Handley, Hobson, and Lasenby}}]{Hergt:2021qlh}
\bibinfo{author}{\bibfnamefont{L.~T.} \bibnamefont{Hergt}}, \bibinfo{author}{\bibfnamefont{W.~J.} \bibnamefont{Handley}}, \bibinfo{author}{\bibfnamefont{M.~P.} \bibnamefont{Hobson}}, \bibnamefont{and} \bibinfo{author}{\bibfnamefont{A.~N.} \bibnamefont{Lasenby}}, \bibinfo{journal}{Phys. Rev. D} \textbf{\bibinfo{volume}{103}}, \bibinfo{pages}{123511} (\bibinfo{year}{2021}), \eprint{2102.11511}.

\bibitem[{\citenamefont{{Particle Data Group} et~al.}(2022)\citenamefont{{Particle Data Group}, Workman, Burkert, Crede, Klempt, Thoma, Tiator, Agashe, Aielli, Allanach et~al.}}]{ParticleDataGroup:2022pth}
\bibinfo{author}{\bibnamefont{{Particle Data Group}}}, \bibinfo{author}{\bibfnamefont{R.~L.} \bibnamefont{Workman}}, \bibinfo{author}{\bibfnamefont{V.~D.} \bibnamefont{Burkert}}, \bibinfo{author}{\bibfnamefont{V.}~\bibnamefont{Crede}}, \bibinfo{author}{\bibfnamefont{E.}~\bibnamefont{Klempt}}, \bibinfo{author}{\bibfnamefont{U.}~\bibnamefont{Thoma}}, \bibinfo{author}{\bibfnamefont{L.}~\bibnamefont{Tiator}}, \bibinfo{author}{\bibfnamefont{K.}~\bibnamefont{Agashe}}, \bibinfo{author}{\bibfnamefont{G.}~\bibnamefont{Aielli}}, \bibinfo{author}{\bibfnamefont{B.~C.} \bibnamefont{Allanach}}, \bibnamefont{et~al.}, \bibinfo{journal}{Progress of Theoretical and Experimental Physics} \textbf{\bibinfo{volume}{2022}}, \bibinfo{pages}{083C01} (\bibinfo{year}{2022}).

\bibitem[{\citenamefont{Feldman and Cousins}(1998)}]{Feldman:1997qc}
\bibinfo{author}{\bibfnamefont{G.~J.} \bibnamefont{Feldman}} \bibnamefont{and} \bibinfo{author}{\bibfnamefont{R.~D.} \bibnamefont{Cousins}}, \bibinfo{journal}{Physical Review D} \textbf{\bibinfo{volume}{57}}, \bibinfo{pages}{3873} (\bibinfo{year}{1998}), ISSN \bibinfo{issn}{0556-2821, 1089-4918}, \eprint{physics/9711021}.

\bibitem[{\citenamefont{Torrado and Lewis}(2020{\natexlab{a}})}]{torradoCobayaCodeBayesian2020}
\bibinfo{author}{\bibfnamefont{J.}~\bibnamefont{Torrado}} \bibnamefont{and} \bibinfo{author}{\bibfnamefont{A.}~\bibnamefont{Lewis}} (\bibinfo{year}{2020}{\natexlab{a}}), \eprint{2005.05290}.

\bibitem[{\citenamefont{James and Roos}(1975)}]{James:1975dr}
\bibinfo{author}{\bibfnamefont{F.}~\bibnamefont{James}} \bibnamefont{and} \bibinfo{author}{\bibfnamefont{M.}~\bibnamefont{Roos}}, \bibinfo{journal}{Comput. Phys. Commun.} \textbf{\bibinfo{volume}{10}}, \bibinfo{pages}{343} (\bibinfo{year}{1975}).

\bibitem[{\citenamefont{Cartis et~al.}(2018)\citenamefont{Cartis, Fiala, Marteau, and Roberts}}]{cartis2018ImprovingFlexibilityRobustnessModelBased}
\bibinfo{author}{\bibfnamefont{C.}~\bibnamefont{Cartis}}, \bibinfo{author}{\bibfnamefont{J.}~\bibnamefont{Fiala}}, \bibinfo{author}{\bibfnamefont{B.}~\bibnamefont{Marteau}}, \bibnamefont{and} \bibinfo{author}{\bibfnamefont{L.}~\bibnamefont{Roberts}} (\bibinfo{year}{2018}), \eprint{1804.00154}.

\bibitem[{\citenamefont{Virtanen et~al.}(2020)\citenamefont{Virtanen, Gommers, Oliphant, Haberland, Reddy, Cournapeau, Burovski, Peterson, Weckesser, Bright et~al.}}]{virtanen2020SciPyFundamentalAlgorithmsscientific}
\bibinfo{author}{\bibfnamefont{P.}~\bibnamefont{Virtanen}}, \bibinfo{author}{\bibfnamefont{R.}~\bibnamefont{Gommers}}, \bibinfo{author}{\bibfnamefont{T.~E.} \bibnamefont{Oliphant}}, \bibinfo{author}{\bibfnamefont{M.}~\bibnamefont{Haberland}}, \bibinfo{author}{\bibfnamefont{T.}~\bibnamefont{Reddy}}, \bibinfo{author}{\bibfnamefont{D.}~\bibnamefont{Cournapeau}}, \bibinfo{author}{\bibfnamefont{E.}~\bibnamefont{Burovski}}, \bibinfo{author}{\bibfnamefont{P.}~\bibnamefont{Peterson}}, \bibinfo{author}{\bibfnamefont{W.}~\bibnamefont{Weckesser}}, \bibinfo{author}{\bibfnamefont{J.}~\bibnamefont{Bright}}, \bibnamefont{et~al.}, \bibinfo{journal}{Nature Methods} \textbf{\bibinfo{volume}{17}}, \bibinfo{pages}{261} (\bibinfo{year}{2020}), ISSN \bibinfo{issn}{1548-7105}, \eprint{1907.10121}.

\bibitem[{\citenamefont{Nygaard et~al.}(2023)\citenamefont{Nygaard, Holm, Hannestad, and Tram}}]{Nygaard_2023}
\bibinfo{author}{\bibfnamefont{A.}~\bibnamefont{Nygaard}}, \bibinfo{author}{\bibfnamefont{E.~B.} \bibnamefont{Holm}}, \bibinfo{author}{\bibfnamefont{S.}~\bibnamefont{Hannestad}}, \bibnamefont{and} \bibinfo{author}{\bibfnamefont{T.}~\bibnamefont{Tram}}, \bibinfo{journal}{Journal of Cosmology and Astroparticle Physics} \textbf{\bibinfo{volume}{2023}}, \bibinfo{pages}{064} (\bibinfo{year}{2023}), ISSN \bibinfo{issn}{1475-7516}.

\bibitem[{\citenamefont{Holm et~al.}(2023)\citenamefont{Holm, Nygaard, Dakin, Hannestad, and Tram}}]{Holm:2023uwa}
\bibinfo{author}{\bibfnamefont{E.~B.} \bibnamefont{Holm}}, \bibinfo{author}{\bibfnamefont{A.}~\bibnamefont{Nygaard}}, \bibinfo{author}{\bibfnamefont{J.}~\bibnamefont{Dakin}}, \bibinfo{author}{\bibfnamefont{S.}~\bibnamefont{Hannestad}}, \bibnamefont{and} \bibinfo{author}{\bibfnamefont{T.}~\bibnamefont{Tram}} (\bibinfo{year}{2023}), \eprint{2312.02972}.

\bibitem[{\citenamefont{Karwal et~al.}(2024)\citenamefont{Karwal, Patel, Bartlett, Poulin, Smith, and Pfeffer}}]{Karwal:2024qpt}
\bibinfo{author}{\bibfnamefont{T.}~\bibnamefont{Karwal}}, \bibinfo{author}{\bibfnamefont{Y.}~\bibnamefont{Patel}}, \bibinfo{author}{\bibfnamefont{A.}~\bibnamefont{Bartlett}}, \bibinfo{author}{\bibfnamefont{V.}~\bibnamefont{Poulin}}, \bibinfo{author}{\bibfnamefont{T.~L.} \bibnamefont{Smith}}, \bibnamefont{and} \bibinfo{author}{\bibfnamefont{D.~N.} \bibnamefont{Pfeffer}} (\bibinfo{year}{2024}), \eprint{2401.14225}.

\bibitem[{\citenamefont{Starobinsky}(1980)}]{Starobinsky}
\bibinfo{author}{\bibfnamefont{A.}~\bibnamefont{Starobinsky}}, \bibinfo{journal}{Physics Letters B} \textbf{\bibinfo{volume}{91}}, \bibinfo{pages}{99} (\bibinfo{year}{1980}).

\bibitem[{\citenamefont{Torrado and Lewis}(2020{\natexlab{b}})}]{Torrado:2020xyz}
\bibinfo{author}{\bibfnamefont{J.}~\bibnamefont{Torrado}} \bibnamefont{and} \bibinfo{author}{\bibfnamefont{A.}~\bibnamefont{Lewis}} (\bibinfo{year}{2020}{\natexlab{b}}), \eprint{2005.05290}.

\bibitem[{\citenamefont{Lewis}(2019)}]{Lewis:2019}
\bibinfo{author}{\bibfnamefont{A.}~\bibnamefont{Lewis}} (\bibinfo{year}{2019}), \eprint{1910.13970}.

\bibitem[{\citenamefont{Hunter}(2007)}]{matplotlib}
\bibinfo{author}{\bibfnamefont{J.~D.} \bibnamefont{Hunter}}, \bibinfo{journal}{Computing in Science \& Engineering} \textbf{\bibinfo{volume}{9}}, \bibinfo{pages}{90} (\bibinfo{year}{2007}).

\bibitem[{\citenamefont{Harris et~al.}(2020)\citenamefont{Harris, Millman, van~der Walt, Gommers, Virtanen, Cournapeau, Wieser, Taylor, Berg, Smith et~al.}}]{numpy}
\bibinfo{author}{\bibfnamefont{C.~R.} \bibnamefont{Harris}}, \bibinfo{author}{\bibfnamefont{K.~J.} \bibnamefont{Millman}}, \bibinfo{author}{\bibfnamefont{S.~J.} \bibnamefont{van~der Walt}}, \bibinfo{author}{\bibfnamefont{R.}~\bibnamefont{Gommers}}, \bibinfo{author}{\bibfnamefont{P.}~\bibnamefont{Virtanen}}, \bibinfo{author}{\bibfnamefont{D.}~\bibnamefont{Cournapeau}}, \bibinfo{author}{\bibfnamefont{E.}~\bibnamefont{Wieser}}, \bibinfo{author}{\bibfnamefont{J.}~\bibnamefont{Taylor}}, \bibinfo{author}{\bibfnamefont{S.}~\bibnamefont{Berg}}, \bibinfo{author}{\bibfnamefont{N.~J.} \bibnamefont{Smith}}, \bibnamefont{et~al.}, \bibinfo{journal}{Nature} \textbf{\bibinfo{volume}{585}}, \bibinfo{pages}{357–362} (\bibinfo{year}{2020}), \eprint{2006.10256}.

\end{thebibliography}

\end{document}